\documentclass[11pt]{article} 
\usepackage[margin=0.75in]{geometry}   
\usepackage{setspace}
\usepackage{tabularx} 
\usepackage{amsmath,amssymb,amsthm} 
\usepackage[square,     
numbers,      
sort&compress 
]{natbib}
\usepackage{graphicx}
\usepackage{float}
\usepackage{mathtools}
\usepackage{mathrsfs} 
\usepackage{varioref}
\usepackage{wrapfig}
\usepackage{threeparttable}
\usepackage{dcolumn}
\newcolumntype{d}{D{.}{.}{-1}}
\usepackage{nomencl}
\makeglossary
\usepackage{subfigure}
\usepackage{subfigmat}
\usepackage{fancyvrb}
\usepackage{ucs}
\usepackage{xcolor}
\usepackage[utf8x]{inputenc}
\usepackage{textgreek}

\begin{document}


\centerline{\LARGE\bf{Effect of Inflow Turbulence on Premixed}}
\medskip
\centerline{\LARGE\bf{Combustion in a Cavity Flameholder}} 
\bigskip
\centerline{\large\it{Gabriel B. Goodwin, Ryan F. Johnson, David A. Kessler, Andrew D. Kercher}}
\medskip
\centerline{Naval Research Laboratory}
\centerline{4555 Overlook Ave. SW, Washington, DC, USA 20375} 
\centerline{Corresponding author: gabe.goodwin@nrl.navy.mil}
\bigskip
\centerline{\large\it{Harsha K. Chelliah}}
\medskip
\centerline{Department of Aerospace Engineering}
\centerline{University of Virginia, Charlottesville, VA}

\doublespacing

\section*{Abstract} 
\let\thefootnote\relax\footnote{DISTRIBUTION A: Approved for public release, distribution is unlimited}
A discontinuous Galerkin finite element method code, JENRE\textsuperscript{\textregistered}, was used to perform highly resolved simulations of ramjet-mode combustion in the University of Virginia Supersonic Combustion Facility cavity flameholder at a flight enthalpy of Mach 5. The primary goal of the work is to enhance our understanding of the effects of turbulence on fully premixed ramjet combustion with a hydrocarbon fuel. Prior experiments measured a freestream turbulence intensity at the inflow to the cavity ranging from 10 -- 15\%. A synthetic turbulence inflow generator was implemented for the simulations in this work to reproduce the turbulence at the inflow to the cavity. This reduced computational expense, as the turbulent, non-reacting flow upstream of the cavity was generated by a boundary condition rather than requiring the modeling of the entire upstream domain. Velocity perturbations and turbulence intensity generated by the turbulent inflow boundary condition are shown to match those values measured in the facility using particle induced velocimetry. Simulations were performed both with and without inflow turbulence to study the effect of turbulence on flame stability and structure. In both cases, a cavity-stabilized flame was achieved. The inflow turbulence promoted more robust combustion, causing the flame to propagate further from the cavity into the core flow, broadening the flame angle with respect to the axial flow direction. The flame angle captured in the simulation agrees with experimental results and theoretical prediction. The cavity's vortex shedding frequency was identified. Flame strain rate and pressure fluctuations in the cavity shear layer were also measured and found to vary periodically at a frequency equal to that of the vortex shedding. High flow strain rate in the cavity shear layer, driven by the vortex shedding process, is identified as a cause of flame stretching and low OH concentrations as the flame traverses the cavity ramp. The effect of spatial resolution on the simulations is discussed through a comparison of cases using second-order and third-order accurate discontinuous Galerkin finite elements.

\noindent
{\sl Keywords:} Dual-mode scramjet; Turbulent combustion; Numerical simulations; Hypersonics; GPU computing



\section{Introduction}

Maintaining flame stability in hydrocarbon-fueled dual-mode scramjet engines operating at flight Mach numbers of 4--6 is a significant challenge, as residence time in the combustor approaches the ignition delay of the fuel-air mixture. A common method for promoting flame stability is to introduce regions of recirculation within the combustor using a cavity flameholder. The shear layer that forms at the leading edge of the cavity results in recirculation of the fuel-air mixture within the cavity, increasing residence time and enhancing combustion completeness \cite{ben2001cavity}. 

Cavity flameholders were found to be successful at stabilizing hydrocarbon flames in supersonic flows, with the stability limits heavily dependent on air inflow conditions, fuel type and injection scheme, and cavity geometry \cite{rasmussen2005stability}. Subsequent experimental work investigated flame stability in cavity flameholders in both ramjet and scramjet modes, observing flame stabilization location to vary with enthalpy of the airflow in both hydrogen and hydrogen/ethylene fuel mixtures \cite{micka2009combustion}. The dynamics governing the location of the flame within the flameholder and spread into the core flow were explored experimentally and numerically \cite{wang2013combustion}. The authors stated that the spread of combustion from the cavity shear layer into the core flow above the cavity was dominated by the traditional diffusion process as well as the convection process associated with the recirculation flows within the cavity, with a strong coupling between the two processes. Axisymmetric cavity flameholders have recently been used to characterize flame stability limits without the interference of corner boundary layer effects found in planar configurations \cite{liu2019cavity}.

The dual-mode, direct-connect scramjet cavity combustor at the University of Virginia Supersonic Combustion Facility (UVASCF), described in detail in \cite{potturi2015large,rockwell2017development,allison2017investigation}, is used to study combustion of hydrocarbon fuels at flight enthalpies up to Mach 5 with a stagnation temperature of 1200 K. In the facility configuration studied in this work, ethylene is injected far upstream of the cavity flameholder such that the flow into the combustor is a relatively homogeneous mixture of ethylene and air. The UVASCF scramjet combustor is capable of operating with a stable, cavity anchored flame in premixed fuel-air mode for long durations with highly repeatable results \cite{rockwell2017development}. Early experimental work in the facility focused on understanding the mean heat release characteristics and dynamic flame behavior \cite{allison2017investigation,geipel2017high}. Fluctuations in heat release were observed to occur primarily in the shear layer between the combustion products in the cavity and the incoming freestream flow of reactants. Recent experiments using planar laser-induced fluorescence (PLIF) have provided detailed measurements and visualizations of ramjet-mode combustion in the UVASCF combustor using an additively manufactured cavity flameholder insert with active cooling of the cavity insert walls \cite{geipel2020AIAAjournal}. Prior experimental configurations at UVASCF used a cavity embedded in the facility wall, similar to that used in earlier cavity flame stabilization research \cite{rasmussen2005stability}. 

\begin{figure}[h!]
	\centering
	\includegraphics[width=67mm]{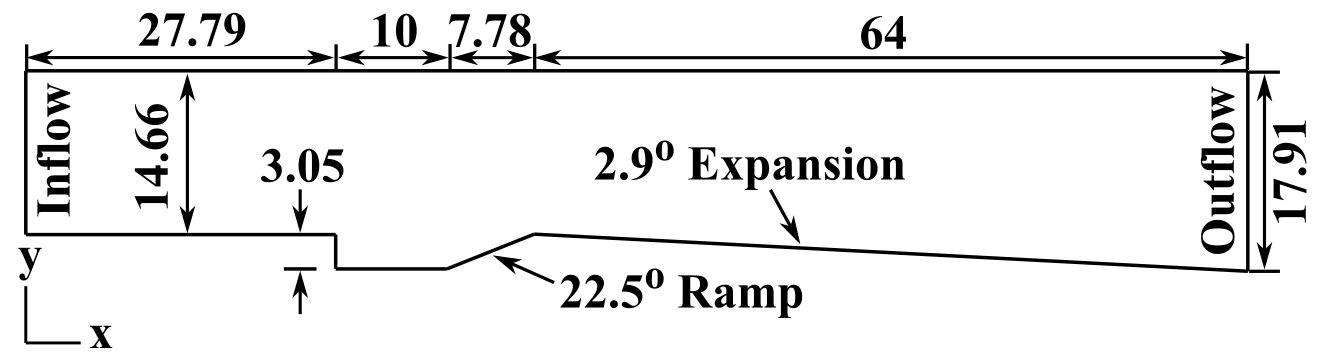}
	\caption{Two-dimensional computational domain with dimensions in mm. Exhaust plenum (radius 1.4 m) not shown.}
	\label{fig:model}
\end{figure}

Hybrid Reynolds-Averaged Navier Stokes/Large Eddy Simulations (RANS/LES) of combustion in the UVASCF cavity with fuel injection upstream of embedded-wall cavity found flame angles captured in the simulation to agree with those predicted by classical premixed turbulent flame speed estimates \cite{potturi2015large,peters1999turbulent}. More recent hybrid RANS/LES computations simulated combustion in the facility in the cavity insert configuration with fuel injection far upstream of the cavity \cite{nielsen2020AIAAjournal}. The simulation results agreed with the aforementioned experimental results using this configuration, available in \cite{geipel2020AIAAjournal}, and provided detailed local flow conditions at the inflow to the cavity flameholder insert. Using this data, boundary conditions were defined for the inflow to the cavity flameholder insert computational domain shown in Fig.~\ref{fig:model}, which is used for the simulations discussed in this work. 

The ability to study the effect of varying levels of freestream turbulence on local fluid dynamics is desirable across a wide range of internal and external, reacting and non-reacting flows. While generating turbulence that is realistic for a given simulation is a challenging problem, there are a number of approaches for synthesizing turbulence at an inflow boundary. Several of these methods were reviewed in recent work by Dhamankar et al. \cite{dhamankar2018overview}, including transition-inducing approaches, turbulence library-based methods, recycling-rescaling based methods, and synthetic turbulence generators. In this work, a synthetic turbulent generator approach is used to produce desired turbulence conditions downstream of the inflow boundary. The specific method used in this work for generating isotropic turbulence at an inflow boundary was introduced by Davidson et al. \cite{Davidson07usingisotropic}.

When performing large eddy simulation (LES) it is often useful to investigate the effect of varying levels of turbulence on the local flow conditions, such as flame stability in a supersonic cavity or boundary layer transition location. Successfully generating the desired turbulence levels in a computationally efficient way is a challenge that has benefited from many years of research, as described in the review by Dhamankar et al. \cite{dhamankar2018overview}. A brief overview of commonly-used methods for generating a turbulent flow with user-defined characteristics is provided here. 

Transition-inducing approaches purposefully trip the laminar flow to turbulent through the use of geometrical features, such as turbulence inducing grids and vanes as described in \cite{shur2014synthetic}, or through the use of artificial surface roughness designed to mimic roughness strips used in experiments, as described in \cite{schlatter2009turbulent}. In a turbulence library-based approach, turbulent data from a prexisting turbulence database from a prior computation or experiment is fed into the main simulation. This method was recently used for the simulation of turbulent combustion in a supersonic cavity \cite{rauch2018dns}. The recycling-rescaling approach by Lund et al. \cite{lund1998generation} uses a simulation running concurrently with the main simulation to develop turbulence and inject it at the inflow boundary to the main simulation. In the concurrent simulation, the inflow velocity field is rescaled from a downstream station and reintroduced at the inflow to the concurrent simulation. Instantaneous velocity profiles are then extracted from the concurrent simulation to be used at the inflow boundary of the main simulation. Finally, the synthetic turbulence generation approach benefits from not requiring a pre-computed library or a concurrent simulation. Instead, the turbulence is generated while the simulation is running in the manner of an inflow velocity profile that fluctuates about a user-defined mean velocity profile. The key challenge in this approach is generating spatially and temporally correlated fluctuations such that realistic turbulent structure develops and propagates downstream without quickly dissipating.

This paper presents the results of highly resolved numerical simulations of the turbulent combustion in the cavity flameholder operating in ramjet mode. Both two and three-dimensional (2D and 3D) simulations were performed. The purpose of the current work is to reproduce experimental results and characterize the effect of turbulence in the incoming flow on flame stability and propagation of the flame into the core flow downstream of the cavity. Reducing computational expense by simulating only the cavity flameholder region was desirable for these simulations, which focus on resolving the fine-scale flame dynamics in the combustor rather than the macroscopic dynamics of the entire facility as was done in previous work \cite{nielsen2020AIAAjournal}. A synthetic turbulence generator was used to reproduce the turbulent flow conditions at the inflow to the cavity, as reported in prior simulations and experiments, and simulations were performed both with and without inflow turbulence. Vortex shedding from the cavity, and the effect on the flame strain rate and pressure fluctuations in the cavity and downstream, is discussed. JENRE\textsuperscript{\textregistered}, the Naval Research Laboratory's discontinuous Galerkin (DG) finite element method code, described in \cite{Joh20JCP}, was used to perform the simulations on a local desktop machine utilizing five Nvidia Tesla V100 graphical processing units (GPUs).  

\section{Numerical Method} \label{DG} \label{numerics}

In this work, we solve a fully conservative formulation of the multi-component, chemically reacting Navier Stokes equations \cite{Joh20JCP}, given as, 

\begin{equation} 
\frac{\partial\rho\boldsymbol{v}}{\partial t}+\nabla\cdot\left(\rho\boldsymbol{v}\otimes\boldsymbol{v}+p\mathbf{\mathbb{I}}\right)=\nabla\cdot\mathbf{\boldsymbol{\tau}},\label{eq:momentum}
\end{equation}

\begin{equation} 
\frac{\partial\left(\rho e_\mathrm{t}\right)}{\partial t}+\nabla\cdot\left(\left(\rho e_\mathrm{t}+p\right)\boldsymbol{v}\right)=\nabla\cdot\left(\lambda\nabla T-\sum_{i=1}^{N_\mathrm{s}}W_\mathrm{i}C_\mathrm{i}h_\mathrm{i}\mathbf{V_\mathrm{i}}+\boldsymbol{\tau}\cdot\boldsymbol{v}\right),\label{eq:total_energy}
\end{equation}

\begin{equation} 
\frac{\partial C_\mathrm{i}}{\partial t}+\nabla\cdot\left(C_\mathrm{i}\left(\boldsymbol{v}+\mathbf{V_\mathrm{i}}\right)\right)=\omega_\mathrm{i}\text{ for }i=1\dots N_\mathrm{s},\label{eq:species_equation}
\end{equation}

\noindent where $\rho$ is density, $\boldsymbol{v}$ is velocity, $p$ is pressure, $\mathbb{I}$ is an identity matrix, $\boldsymbol{\tau}$ is the deviatoric stress tensor as defined in Eq. (2.17) in \cite{Joh20JCP}, $e_\mathrm{t}$ is total energy, $\lambda$ is conductivity, $T$ is temperature, $N_\mathrm{s}$ is the number of species, $W_\mathrm{i}$ is molecular weight, $C_\mathrm{i}$ is concentration, $h_\mathrm{i}$ is enthalpy, $\mathbf{V}_\mathrm{i}$ is diffusion velocity, $\omega_\mathrm{i}$ is the production source term, and subscript $i$ indicates species $i$. The production source term is calculated from the progress reaction rates for any number of reactions and reaction types. Pressure is given by the equation of state, 

\begin{equation}
p=R^{o}T\sum_{i=1}^{N_\mathrm{s}}C_\mathrm{i}\label{eq:EOS},
\end{equation}

\noindent where $R^o$ is the universal gas constant. The species diffusion velocity is given by

\begin{equation}
\mathbf{V_\mathrm{i}}=\frac{D_\mathrm{i}}{C_\mathrm{i}}\nabla C_\mathrm{i}-\frac{D_\mathrm{i}}{\rho}\nabla\rho,\label{eq:diffusion_velocity}
\end{equation}

\noindent where $D_\mathrm{i}$ is the mass-averaged diffusion coefficient. All transport coefficients are calculated using mixture-averaged approaches \cite{Wil50,Mat67,Kee89}. As demonstrated in \cite{Joh20JCP}, no artificial viscosity, stabilization, or filtering methods were required to stabilize the multi-component, chemically reacting flows simulated in this work.

In this model, the total energy is defined as

\begin{equation}
\rho e_\mathrm{t}=\rho u+\frac{1}{2}\rho\boldsymbol{v}\cdot\boldsymbol{v},\label{eq:total_energy-1}
\end{equation}

\noindent where $u$ is total energy. Internal energy can be defined as a function of total enthalpy, $\rho u=\rho h_\mathrm{t}-p$,
where $h_\mathrm{t}$ is total enthalpy and $\rho h_\mathrm{t}=\sum_{i=1}^{N_\mathrm{s}}W_\mathrm{i}C_\mathrm{i}h_\mathrm{i}$, where $h_i$ is a nonlinear function with respect to temperature. 

Equations (\ref{eq:momentum}--\ref{eq:species_equation}) are discretized using the DG method. The resulting DG space semi-discretization, Eq. (3.6) of \cite{Joh20JCP}, is integrated temporally with a second order strong-stability-preserving Runge-Kutta method \cite{Got01}.  The temporal integration of the source term is separated from the temporal integration of the conservation laws via Strang operator splitting.  The resulting system of ordinary differential equations, describing the influence of the source term on the temporal evolution of the state, is integrated using a temporal and polynomial basis adaptive DG method, DGODE \cite{Joh20JCP}. A 19 species, 35 reaction step elementary skeletal mechanism is used to model the ethylene-air combustion \cite{dong2008numerical}. 

\section{Computational Geometry \& Problem Set-Up} \label{experiment}


The computational domain is shown in Fig.~\ref{fig:model}. This domain is a 2D slice through the planar cavity flameholder insert. The domain used for the 3D simulations is identical to that shown in Fig.~\ref{fig:model}, but extruded in the $z$-direction. The plenum into which the combustor outflow exhausts is not shown in Fig.~\ref{fig:model}; it is 1.4 m in radius with atmospheric pressure fixed at the outflow boundary and remaining variables interpolated from the interior. The plenum was used in order to minimize downstream boundary condition influence on the flow in the combustor. 

The following discusses the computational mesh used for the 2D simulations; the 3D simulations are discussed in detail in Sec.~\ref{3D}. The Kolmogorov length scale in the cavity shear layer and flame was calculated to be approximately 10 $\mathrm{\mu}$m and the width of the smallest flame structures observed in the experiment using high-spatial-resolution OH PLIF was 110 $\mathrm{\mu}$m \cite{geipel2017high,geipel2020AIAAjournal}. This was used to guide the computational mesh development for the present study. The computational element size for the 2D simulations ranges from 30 $\mathrm{\mu}$m in the cavity shear layer, flame, and against the walls to 220 $\mathrm{\mu}$m in the core flow. Prior work has shown that resolution for DG simulations can be approximated to $h/(p+1)$ where $h$ is the mesh element size and $p$ is the order of accuracy of the DG element \cite{moura2017eddy}. It has also been shown that choice of DG integration and nodal basis can greatly improve results for a given $h$ \cite{gassner2013accuracy}. For the initial set of 2D simulations in this study, described in Sec.~\ref{laminar} - \ref{pressure_fluct}, second-order accurate DG($p=1$) triangular mesh elements were used, providing an approximate resolution of 15 $\mathrm{\mu}$m in cavity shear layer and flame and 110 $\mathrm{\mu}$m in the core flow. Third-order accurate DG($p=2$) elements were used in a subsequent 2D simulation described in Sec.~\ref{high_order} and it is shown that the DG($p=1$) elements are sufficient for resolving all flame structures.

\begin{figure}[h!]
	\centering
	\includegraphics[width=144mm]{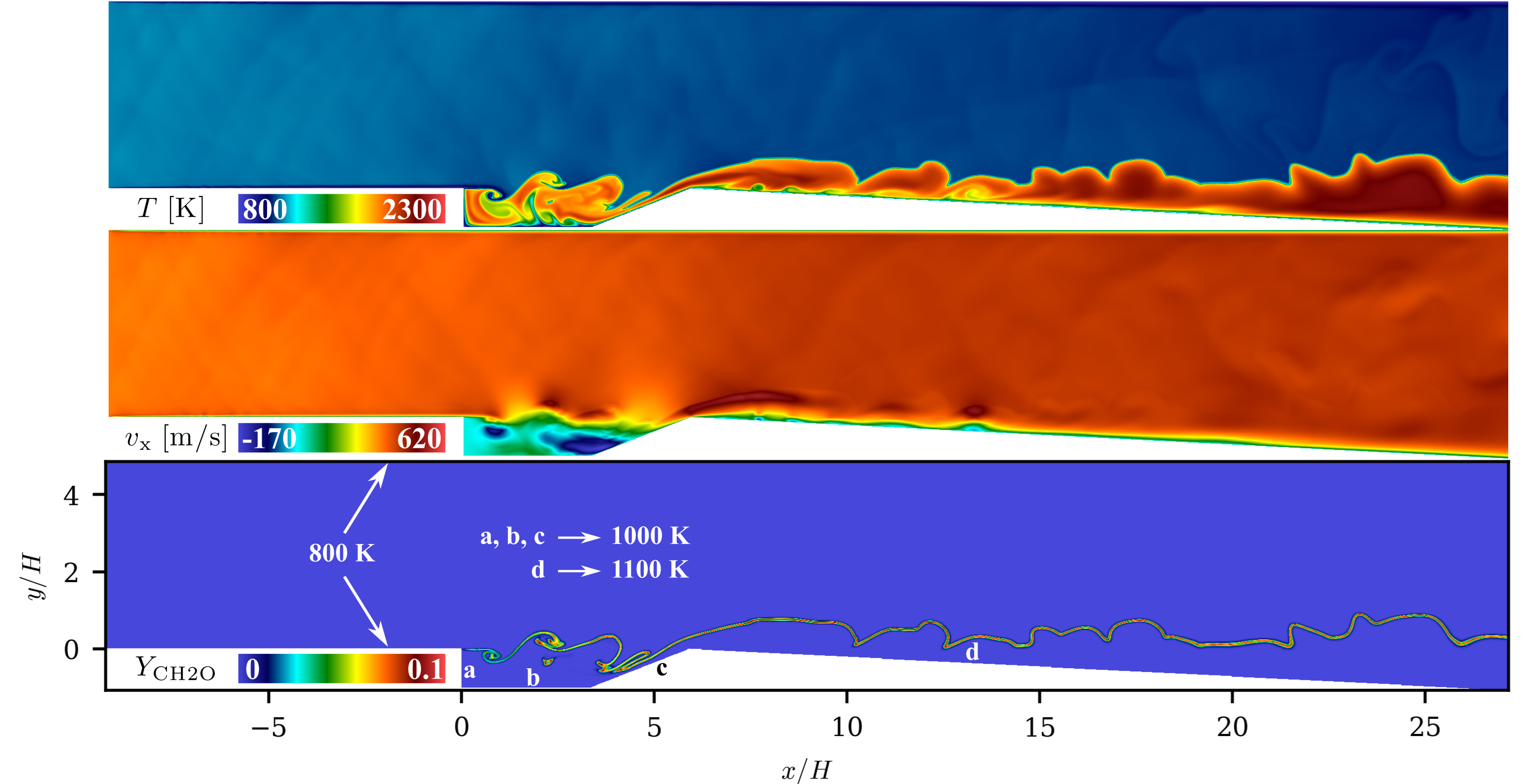}
	\caption{Contours of $T$, $v_\mathrm{x}$, and $Y_\mathrm{CH2O}$ show a cavity-stabilized flame with a laminar inflow. Isothermal wall temperatures are shown.}
	\label{fig:laminar_flame}
\end{figure}

The left boundary is a subsonic inflow of a homogeneous ethylene-air mixture at Mach 0.6, 1.72 atm, and 1125 K, with a fuel equivalence ratio, $\phi$, of 0.6. Simulations were performed with both laminar and turbulent inflows. The wall boundary conditions are no-slip and isothermal, with wall temperatures set to approximate the steady-state wall temperature in the experimental facility. Experimental wall temperature data for the actively-cooled cavity insert is not available, so the wall temperatures were obtained from prior simulations \cite{nielsen2020AIAAjournal}. The temperature conditions for the isothermal walls are shown in Fig.~\ref{fig:laminar_flame}. 

\section{Results \& Discussion} \label{results}

\subsection{Two-Dimensional Simulations}

\subsubsection{Combustion with Laminar Inflow} \label{laminar}

Figure~\ref{fig:laminar_flame} shows a cavity-stabilized flame with a laminar inflow boundary condition with $x$ and $y$ axes normalized by the cavity height, $H$, of 3.05 mm. The ethylene-air mixture was ignited after two flow through residence times through the combustor by initializing a circular high-temperature region in the cavity. The flame quickly stabilized and anchored to the cavity leading edge. The $T$, $x$-velocity, $v_\mathrm{x}$, and mass fraction of formaldehyde, $Y_\mathrm{CH2O}$, contours shown in Fig.~\ref{fig:laminar_flame} are sampled two flow through residence times following ignition. There is a roll-up of the flame immediately downstream of the cavity leading edge due to interaction of the flame with the cavity shear layer. In the expanding section, the flame lifts from the lower wall and propagates into the core flow. Two distinct recirculation zones are visible in the $v_\mathrm{x}$ plot, with a minimum $v_\mathrm{x}$ of -250 m/s at the cavity ramp wall for the present laminar inflow case. The peak $Y_\mathrm{CH2O}$ contour tracks the flame surface, indicating a wrinkled flame front. 

\subsubsection{Synthetic Turbulence Generation} \label{turb_inflow}

A synthetic turbulence inflow boundary condition was implemented to generate turbulence at the inflow that reproduces the intensity measured in experiments at UVASCF. The goal in using this boundary condition was to be able to explore the effect of turbulence on the flame without simulating the entire domain upstream of the cavity flameholder, using a pre-computed turbulence library, or running a concurrent simulation to generate the turbulence. The method of generating synthetic turbulence at an inflow boundary condition from Davidson et al. was selected \cite{Davidson07usingisotropic}. This section describes the procedure for using this method to generate spatially and temporally correlated turbulent velocity fluctuations about a mean velocity profile.

\begin{figure}[h!]
	\centering
	\includegraphics[width=82mm]{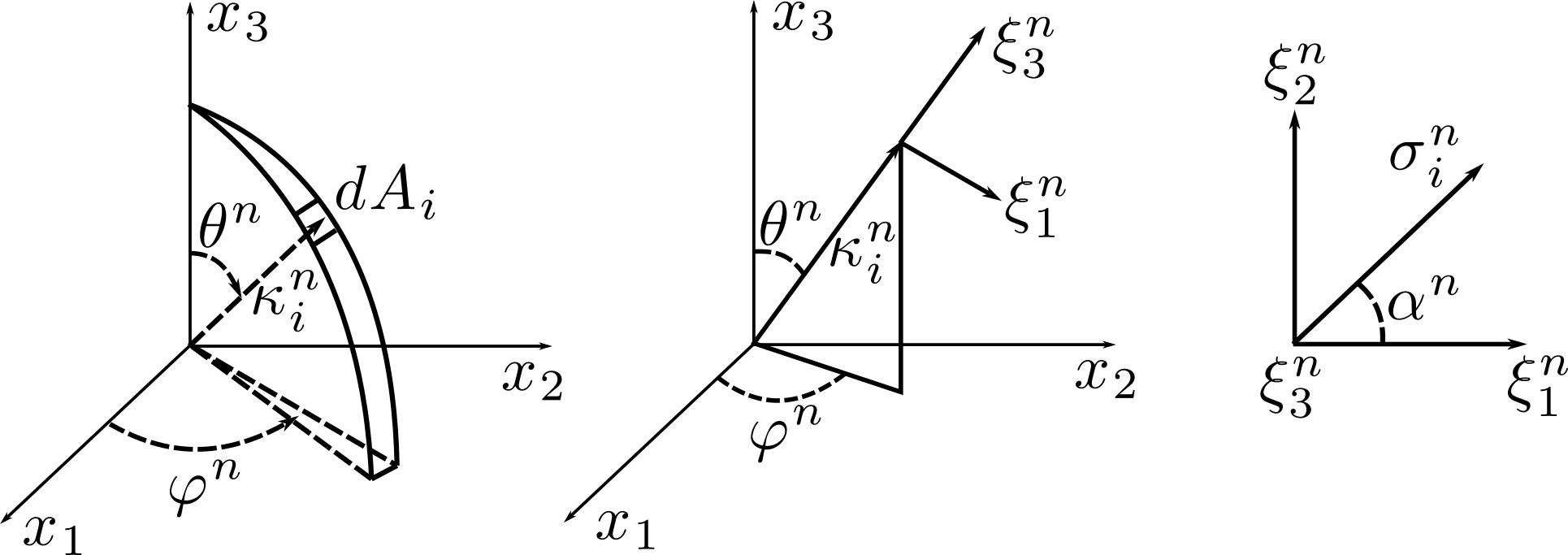}
	\caption{Visualization of random angles $\alpha^n$, $\theta^n$, and $\varphi^n$ and their relation to wave number vector $\kappa^n_i$ and velocity unit vector $\sigma^n_i$.}
	\label{fig:random_angles}
\end{figure}

This method generates isotropic fluctuations by decomposing the turbulent signal into $N$ Fourier modes:

\begin{equation}
v'_i(x_j) = 2 \sum_{n=1}^{N}\hat{u}^n \mathrm{cos}(\kappa^n_j x_j + \Psi^n)\sigma^n_i,\label{eq:turbulent_signal}
\end{equation}

\noindent where $v'$ is the velocity fluctuation, $x$ is the spatial direction, $\hat{u}^n$ is the amplitude of the fluctuation, $\Psi^n$ is the phase of the fluctuation, $\sigma_i^n$ is the direction of Fourier mode $n$, and $k_j^n$ is the wave number for mode $n$. Angles $\varphi^n$, $\alpha^n$, $\theta^n$, and phase $\Psi^n$ are randomly selected at every Fourier mode with the probability distributions given in Table~\ref{table:probabilities}. Figure~\ref{fig:random_angles} shows the orientation of these angles in wave-space and physical space.

Next, a finite region of wavenumber space is chosen and discretized between $\kappa_l$ and $\kappa_\mathrm{max}$, where

\begin{equation}
\kappa_\mathrm{max} = \frac{2\pi}{2\Delta}, \; \kappa_l = \frac{\kappa_e}{p}. 
\end{equation}

\begin{figure}[h!]
	\centering
	\includegraphics[width=83mm]{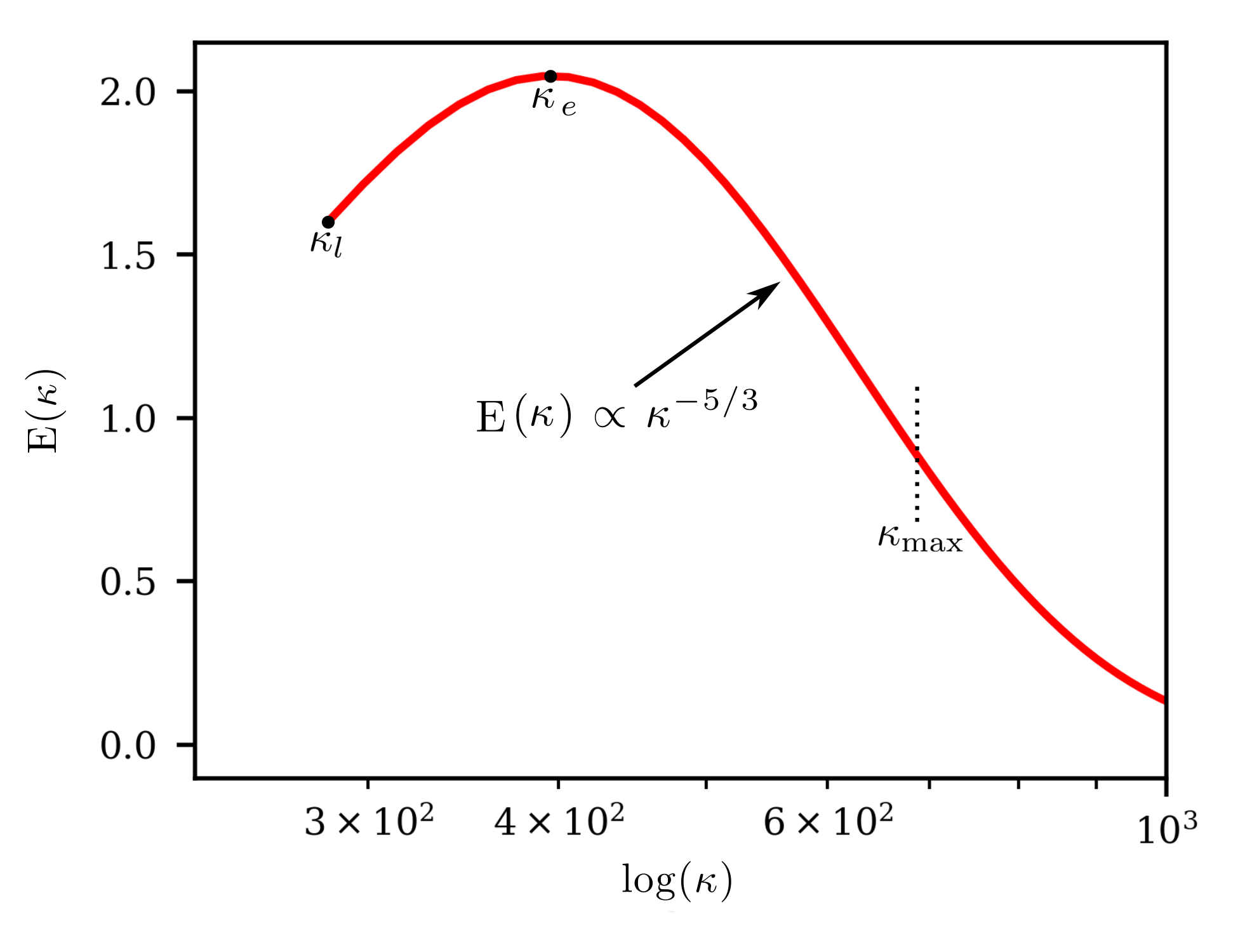}
	\caption{Energy cascade obeys a modified von Karman spectrum.}
	\label{fig:energy}
\end{figure}

\noindent Following Davidson \cite{Davidson07usingisotropic}, the wavenumber associated with the highest energy content is approximated by $\kappa_e = \alpha\frac{9\pi}{55l_t}$ where $\alpha$ is a constant (here set to 1.453), $l_t$ is the turbulent length scale, $\Delta$ is the grid spacing, and $p$ is a constant $>$1 to make the largest scales larger than those corresponding to $\kappa_e$. The wave number space, $\kappa_\mathrm{max}$ - $\kappa_l$, is divided into $N$ modes with equal size $\Delta\kappa$. The components of the wave number vector $\kappa_i^n$ are calculated in accordance with Table~\ref{table:wave_number_vector} and the components of the unit number vector $\sigma_i^n$ are calculated in accordance with Table~\ref{table:unit_vector}. The amplitude is calculated as 

\begin{equation}
\hat{u}^n = \sqrt{E(|\kappa_j^n|)\Delta\kappa},
\end{equation}

\noindent where energy $E(\kappa)$ obeys a modified von Karman spectrum, defined as

\begin{equation}
E(\kappa) = \alpha\frac{v^2_{rms}}{\kappa_e}\frac{(\kappa/\kappa_e)^4}{[1 + (\kappa/\kappa_e)^2]^{17/6}}e^{-2(\kappa/\kappa_\eta)^2},
\end{equation}

\noindent where $\kappa = (\kappa_i \kappa_i)^{1/2}$, $\kappa_\eta = \epsilon^{1/4}\nu^{-3/4}$, and $v_{rms}$ is the root mean square of the velocity, calculated as $v_{rms} = (\frac{2}{3}\kappa)^{1/2}$. An example of an energy spectrum generated by the boundary condition, using the input parameters in Table~\ref{table:parameters}, is shown in Fig.~\ref{fig:energy}. In order to correlate the fluctuations calculated for each independent time step, the new velocity fluctuation is calculated for time step $m$ as

\begin{equation}
(V'_{i})^m = a(V'_{i})^{m-1} + b(v'_{i})^m,
\end{equation}

\noindent where $a = e^{-\Delta t/ \Upsilon}$, $b = \sqrt{1 - a^2}$, $\Upsilon$ is the inflow time scale, and $\Delta t$ is the time step.

\begin{table}
	\begin{center}
		\begin{tabular}{|c|l|} 
			\cline{1-2} 
			$p(\varphi^n) = 1/2\pi$ & $0 \leq \varphi^n \leq 2\pi$  \tabularnewline
			\cline{1-2}
			$p(\Psi^n) = 1/2\pi$ & $0 \leq \Psi^n \leq 2\pi$ \tabularnewline
			\cline{1-2}
			$p(\theta^n) = 1/2\mathrm{sin}(\theta)$ & $0 \leq \theta^n \leq \pi$ \tabularnewline
			\cline{1-2}
			$p(\alpha^n) = 1/2\pi$ & $0 \leq \alpha^n \leq 2\pi$ \tabularnewline
			\cline{1-2}
		\end{tabular}
		\caption{Probability distributions.}
		\label{table:probabilities} 
	\end{center}
\end{table}

\begin{table}
	\begin{center}
		\begin{tabular}{|c|l|} 
			\cline{1-2} 
			$\kappa_{x1}^n$ & $\mathrm{sin}(\theta^n)\mathrm{cos}(\varphi^n)$  \tabularnewline
			\cline{1-2}
			$\kappa_{x2}^n$ & $\mathrm{sin}(\theta^n)\mathrm{sin}(\varphi^n)$ \tabularnewline
			\cline{1-2}
			$\kappa_{x3}^n$ & $\mathrm{cos}(\theta^n)$\tabularnewline
			\cline{1-2}
		\end{tabular}
		\caption{Wave number vector components.}
		\label{table:wave_number_vector} 
	\end{center}
\end{table}

\begin{table}
	\begin{center}
		\begin{tabular}{|c|l|} 
			\cline{1-2} 
			$\sigma_{x1}^n$ & $\mathrm{cos}(\varphi^n)\mathrm{cos}(\theta^n)\mathrm{cos}(\alpha^n) - \mathrm{sin}(\varphi^n)\mathrm{sin}(\alpha^n)$  \tabularnewline
			\cline{1-2}
			$\sigma_{x2}^n$ & $\mathrm{sin}(\varphi^n)\mathrm{cos}(\theta^n)\mathrm{cos}(\alpha^n) + \mathrm{cos}(\varphi^n)\mathrm{sin}(\alpha^n)$ \tabularnewline
			\cline{1-2}
			$\sigma_{x3}^n$ & $-\mathrm{sin}(\theta^n)\mathrm{cos}(\alpha^n)$ \tabularnewline
			\cline{1-2}
		\end{tabular}
		\caption{Unit vector components.}
		\label{table:unit_vector} 
	\end{center}
\end{table}

\begin{table}
	\begin{center}
		\begin{tabular}{|c|l|} 
			\cline{1-2} 
			Input Parameter & Value  \tabularnewline
			\cline{1-2}
			$\Delta$ & 200 $\mathrm{\mu}$m  \tabularnewline
			\cline{1-2}
			$\alpha$ & 1.453  \tabularnewline
			\cline{1-2}
			$p$ & 1.3  \tabularnewline
			\cline{1-2}
			$L_t$ & 7 mm  \tabularnewline
			\cline{1-2}
			$\epsilon$ & 0.425  \tabularnewline
			\cline{1-2}
			$\nu$ & 1 x 10$^{-4}$  \tabularnewline
			\cline{1-2}
			$v_{rms}$ & 40 m/s  \tabularnewline
			\cline{1-2}
			$\Upsilon$ & 1.57075 x 10$^{-7}$ s  \tabularnewline
			\cline{1-2}
		\end{tabular}
		\caption{Synthetic turbulent inflow input parameters.}
		\label{table:parameters} 
	\end{center}
\end{table}

\begin{figure}[h!]
	\centering
	\includegraphics[width=67mm]{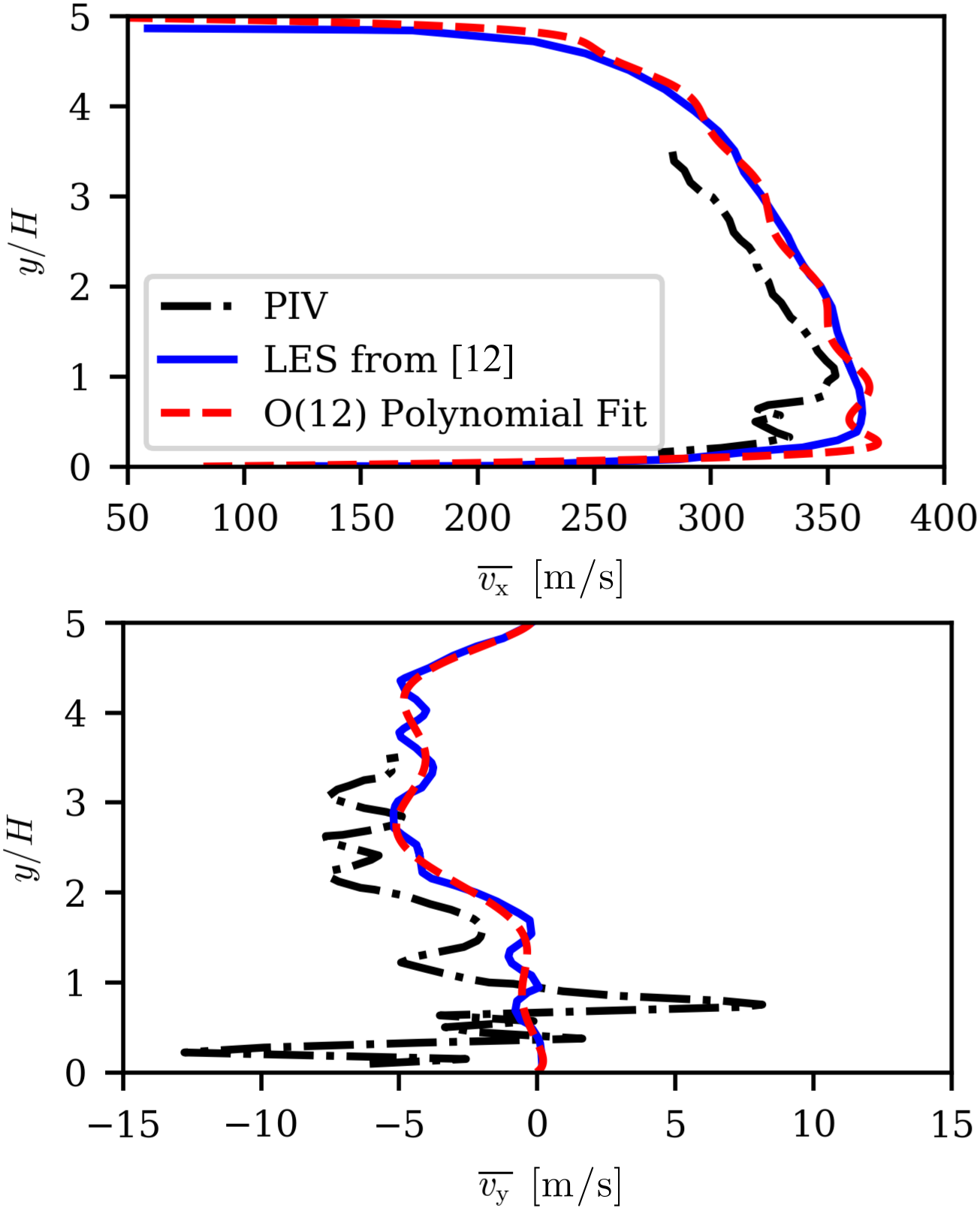}
	\caption{$\overline{v_\mathrm{x}}$ and $\overline{v_\mathrm{y}}$ profiles across $x/H$ = -0.05 used for turbulent inflow.}
	\label{fig:exp_lesv3_compare}
\end{figure}

This method generates isotropic turbulence at the inflow boundary by inducing spatially and temporally correlated velocity perturbations about a user-defined mean velocity, $\overline{\boldsymbol{v}}$, profile. In the hybrid RANS/LES calculations discussed in \cite{nielsen2020AIAAjournal}, the computed $\overline{\boldsymbol{v}}$ profiles are provided at $x/H$ = -0.05 and show good agreement with the particle induced velocimetry (PIV) data for the portion of the cavity height where PIV data is available. These profiles are shown in Fig.~\ref{fig:exp_lesv3_compare}. Due to the sizing of the optical window in the combustor at UVASCF, the PIV measurements did not extend to the top wall of the combustor. Twelfth-degree polynomials were fit to the LES velocity profiles, as shown in Fig.~\ref{fig:exp_lesv3_compare}, and taken as the user-defined mean $x$ and $y$-velocity profiles ($\overline{v_\mathrm{x}}$ and $\overline{v_\mathrm{y}}$, respectively) for the turbulent inflow boundary condition.

\begin{figure}[h!]
	\centering
	\includegraphics[width=144mm]{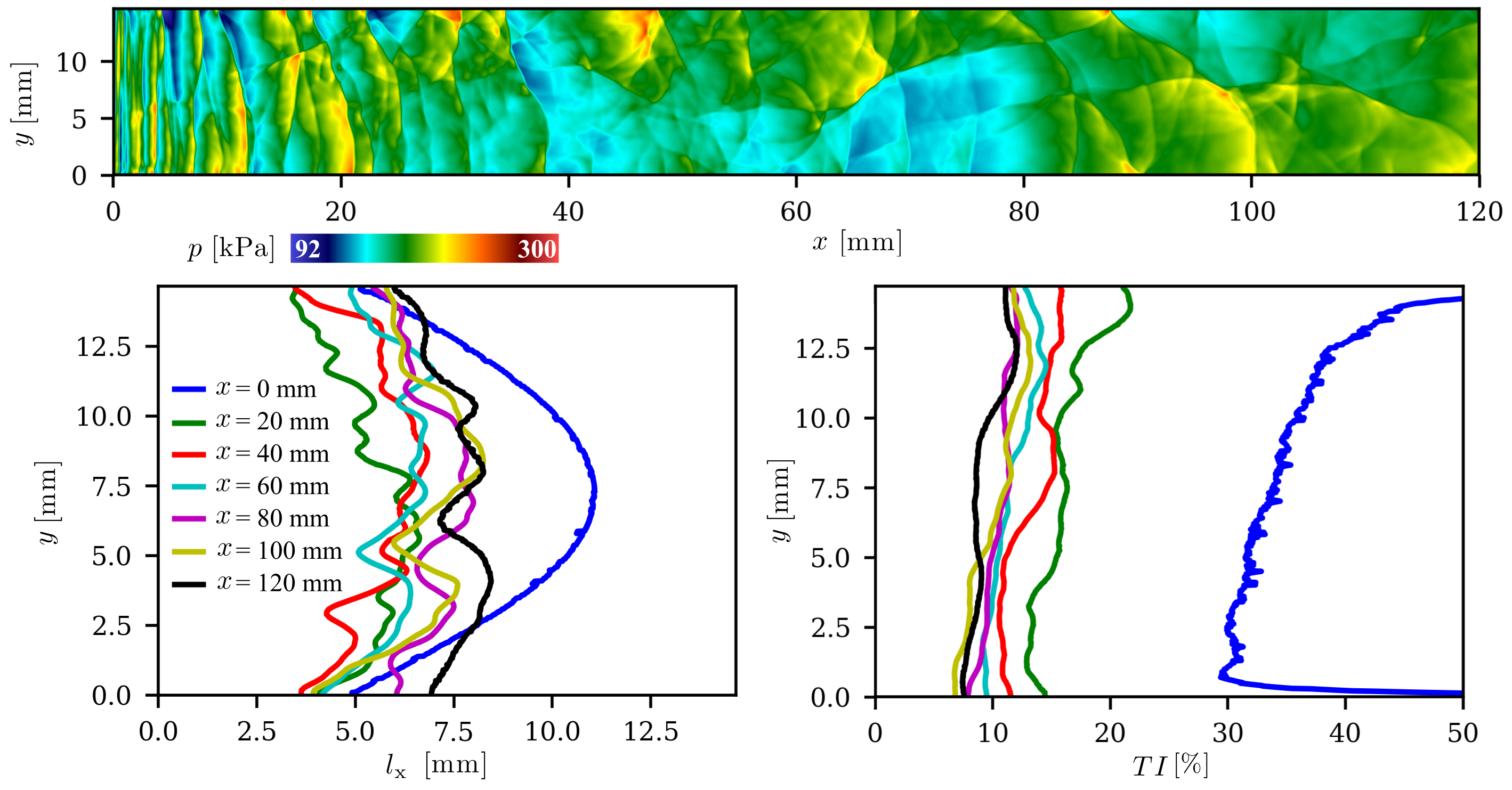}
	\caption{Turbulent inflow into a constant-area channel with $l_\mathrm{x}$ and $TI$ plotted across several vertical slices through domain.}
	\label{fig:turb_inflow_test}
\end{figure}

A simulation was performed to test the turbulent inflow boundary condition and measure the turbulence intensity, $TI$, and turbulent length scale in a constant-area 120 mm long channel, with channel height equal to that of the inlet in the cavity simulations, 14.66 mm. The turbulent inflow was applied at the left boundary (input parameters from Table~\ref{table:parameters}), the top and bottom boundaries were slip walls, and the right boundary was a transmissive outflow. Figure~\ref{fig:turb_inflow_test} shows pressure contours in the domain after two flow through residence times as well as $TI$ and turbulent length scale in the streamwise direction, $l_\mathrm{x}$, across several vertical cross sections. Turbulence intensity decays downstream of the boundary condition and approaches an asymptotic value of $\sim$10\%. Turbulent length scale, calculated using the two-point correlation method outlined in \cite{Davidson07usingisotropic}, also settles from its forced profile at the inflow boundary to an asymptotic value of $\sim$7.5 mm, approximately half the channel height. These results indicate that the turbulent inflow boundary condition generates spatially correlated turbulence with intensity and length scale that settle to stable values downstream of their forced condition at the inflow boundary.

\begin{figure}[h!]
	\centering
	\includegraphics[width=144mm]{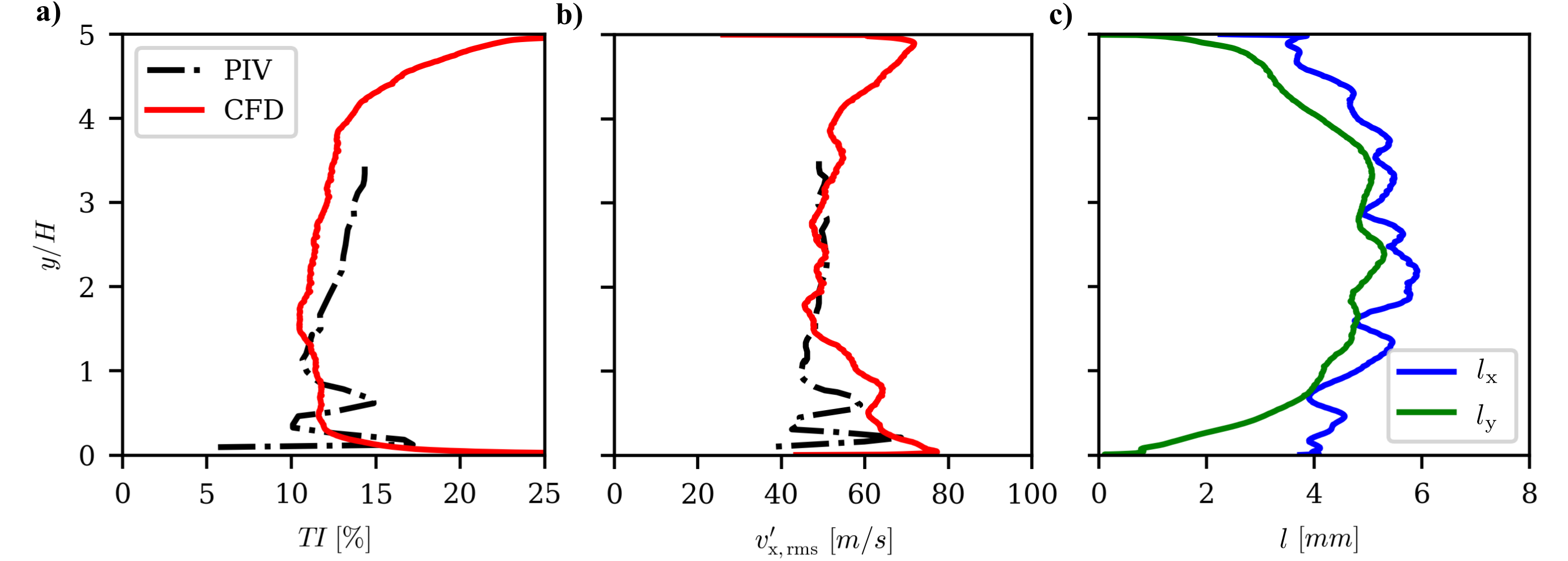}
	\caption{(a) $TI$ (b) $v'_\mathrm{x,rms}$ (c) $l_\mathrm{x}$ and $l_\mathrm{y}$ across a vertical slice through the combustor at $x/H$ = -0.05.}
	\label{fig:exp_compare}
\end{figure}

\subsubsection{Combustion with Turbulent Inflow} \label{turb}

The laminar inflow solution shown in Fig.~\ref{fig:laminar_flame} was used as the initial condition for a second simulation in which the synthetic turbulence boundary condition was applied to the cavity inflow. The prescribed conditions for the turbulent inflow were the same as were used in the constant-area channel test case. Figure~\ref{fig:exp_compare} shows $TI$, root-mean-square of the $x$-velocity perturbations, $v'_\mathrm{x,rms}$, turbulent length scale in the spanwise direction, $l_\mathrm{y}$, and $l_\mathrm{x}$ plotted across a vertical slice through the domain at $x/H$ = -0.05. The data was sampled for two flow through residence times after starting the simulation with the turbulent inflow. The PIV measurements from UVASCF, also taken at $x/H$ = -0.05, are shown. The computational results agree well with the PIV data, with both showing a $TI$ ranging from 10 to 15\% in the core flow. Turbulence intensity is more variable in the experimental data close to the wall, whereas in the simulation it increases steadily approaching the wall. In both the experiments and computations, the $v'_\mathrm{x,rms}$ is approximately 45 m/s in the core flow, increasing slightly near the lower wall. The turbulent length scales in both streamwise and spanwise directions are approximately 5 mm in the core flow. Similarity in $l_\mathrm{x}$ and $l_\mathrm{y}$ is expected, as the turbulence generated at the inflow is isotropic. 

\begin{figure}[h!]
	\centering
	\includegraphics[width=144mm]{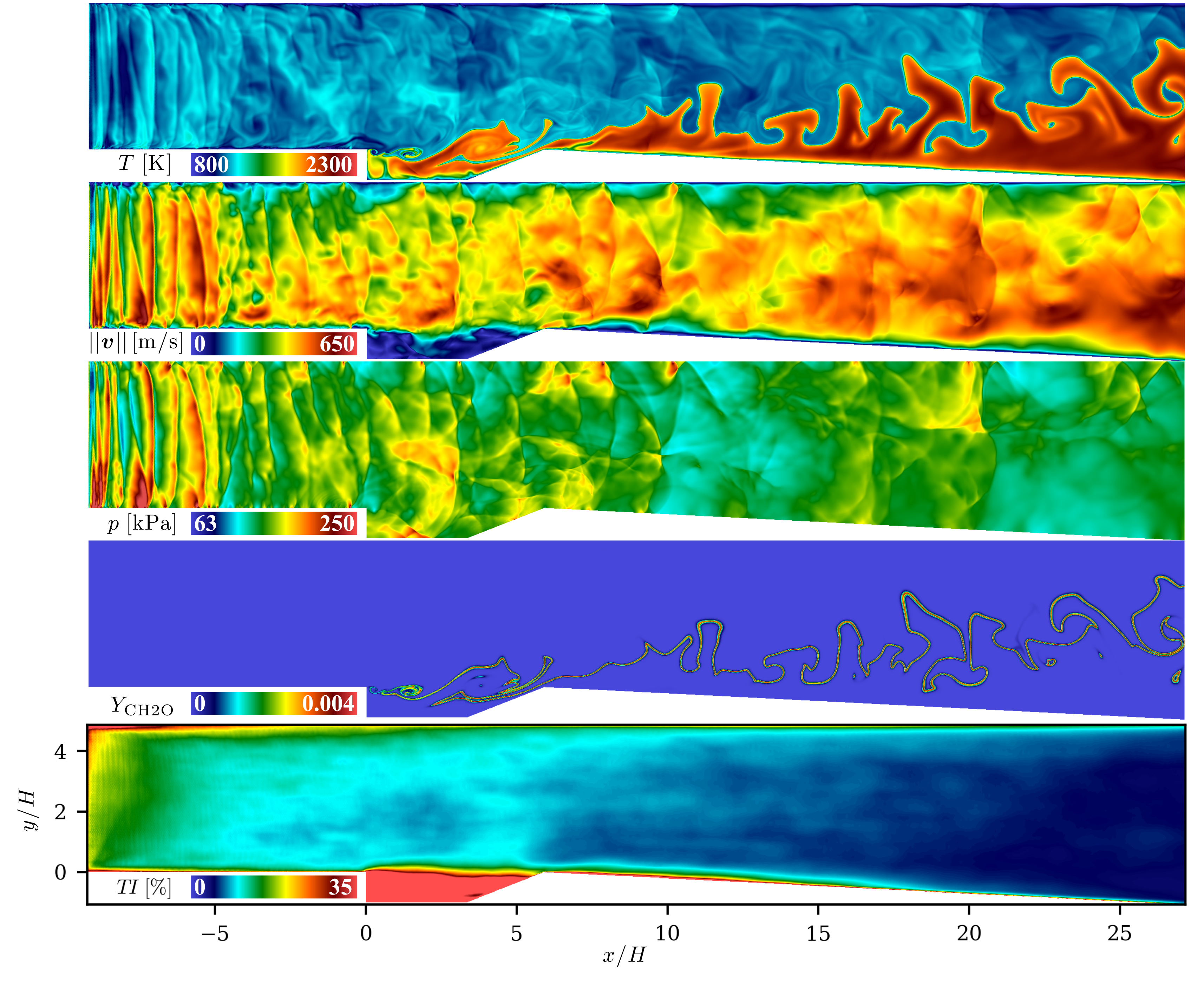}
	\caption{Contours of $T$, $||\boldsymbol{v}||$, $p$, $Y_\mathrm{CH2O}$, and $TI$ for a cavity-stabilized flame with a turbulent inflow.}
	\label{fig:turbulent_flame}
\end{figure}

Figure~\ref{fig:turbulent_flame} shows contours of $T$, velocity magnitude, $||\boldsymbol{v}||$, $p$, and $Y_\mathrm{CH2O}$ sampled two flow through residence times after initializing the simulation. The velocity perturbations generated at the inflow are observed in the $||\boldsymbol{v}||$ plot. As in the constant-area channel test case, the high-frequency waves gradually diffuse as they propagate farther into the domain. The pressure waves collide with and perturb the flame, causing the formation of turbulent structure along the flame surface. Despite the distortion to the flame surface, the flame remains anchored to the cavity leading edge indefinitely, consistent with experiments. Downstream of the cavity, in the expanding section of the combustor, the flame travels further into the core flow than in the laminar inflow case, resulting in higher fuel consumption over a shorter axial distance. Figure~\ref{fig:turbulent_flame} also shows contours of time-averaged $TI$. The $TI$ is $\sim$20\% in the core flow immediately downstream of the inflow boundary, gradually decaying to a value of 10\% in the core flow just upstream of the cavity leading edge; this was the targeted value to match experimental conditions as shown in Fig.~\ref{fig:exp_compare}a. Recirculation in the cavity and near the walls causes $TI$ to remain higher in these regions than in the core flow.

\begin{figure}[h!]
	\centering
	\includegraphics[width=67mm]{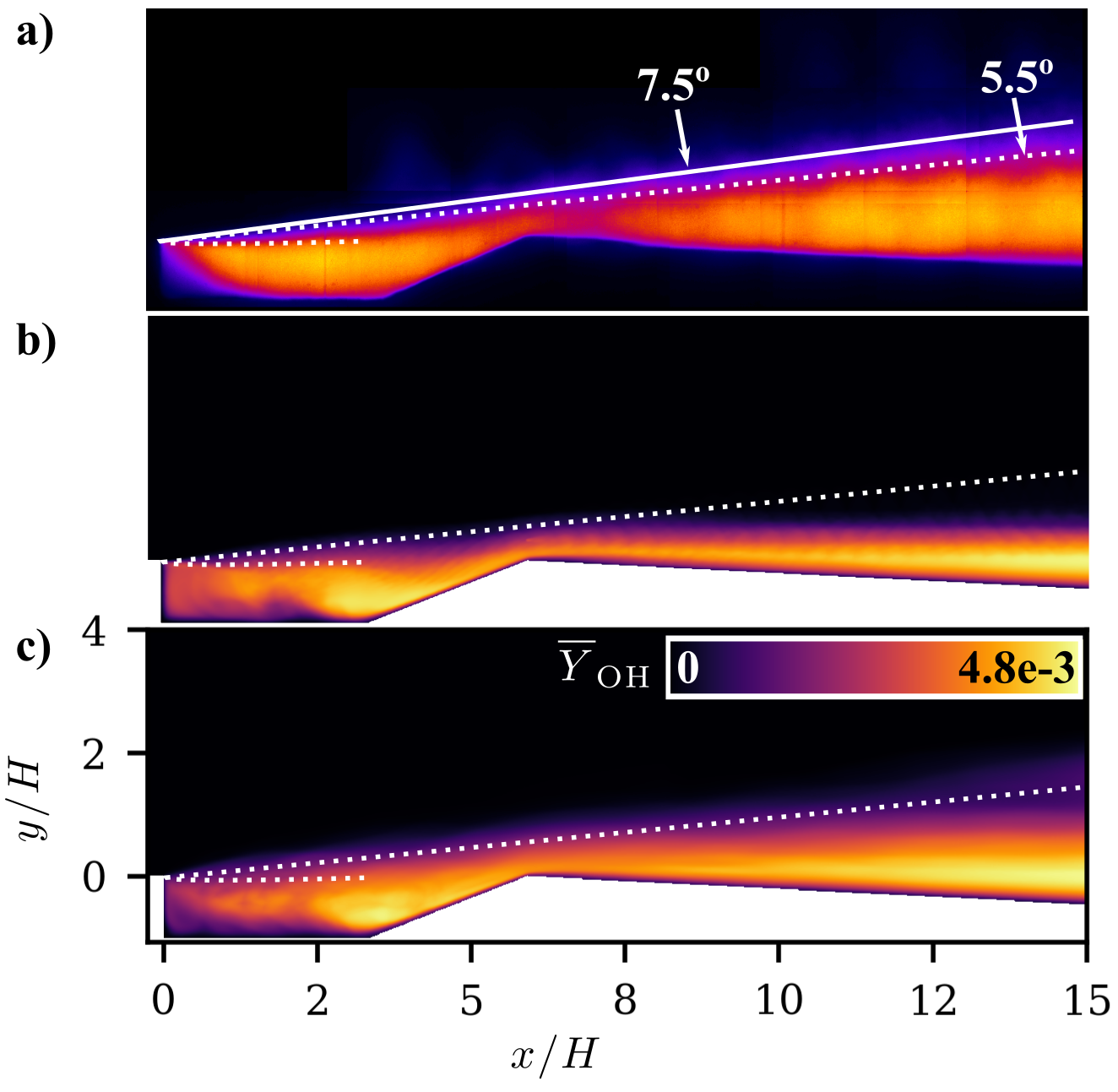}
	\caption{(a) Mean OH PLIF signal from UVASCF (b) $\overline{Y}_\mathrm{OH}$ from laminar inflow simulation (c) $\overline{Y}_\mathrm{OH}$ from tubulent inflow simulation.}
	\label{fig:oh_comparison}
\end{figure}

Figure~\ref{fig:oh_comparison} shows the time-averaged hydroxyl radical mass fraction, $\overline{Y}_\mathrm{OH}$, for the laminar and turbulent inflow simulations and the mean OH PLIF signal from the experiment, all plotted over the same spatial domain. In both simulation and experiment, there is a decrease in OH concentration over the intersection of the cavity ramp and wall of the expanding section (at $y/H$ = 0 and $x/H$ = 6.5). The sharpness of the turning angle at this location results in local stretching of the flame and perhaps some extinguished flame structures, resulting in less OH production. The flow recovers immediately downstream. This localized stretching of the flame as it traverses the cavity ramp was captured in the simulations both with and without a turbulent inflow (see Figs.~\ref{fig:laminar_flame} and~\ref{fig:turbulent_flame}) and was also reported in the experiments \cite{geipel2020AIAAjournal}. The turbulent flame speed correlation of Peters \cite{peters1999turbulent} was used to calculate the theoretical flame angle emanating from the cavity leading edge, shown as the solid white line in Fig.~\ref{fig:oh_comparison}a. The experimentally measured flame angle, marked with the white dashed line, is slightly shallower than the theoretical prediction. The laminar inflow simulation did not reproduce the experimental flame angle, with all OH formation occurring much closer to the bottom wall than in the experiment. When the turbulent inflow was used the flame angle shows good agreement with the experiment indicating that the 2D simulation accurately captures the time-averaged flame location. This result highlights the importance of freestream turbulence in promoting propagation of the flame into the incoming flow of reactants. 

\subsubsection{Vorticity in the Cavity Shear Layer}

\begin{figure}
	\centering
	\includegraphics[width=144mm]{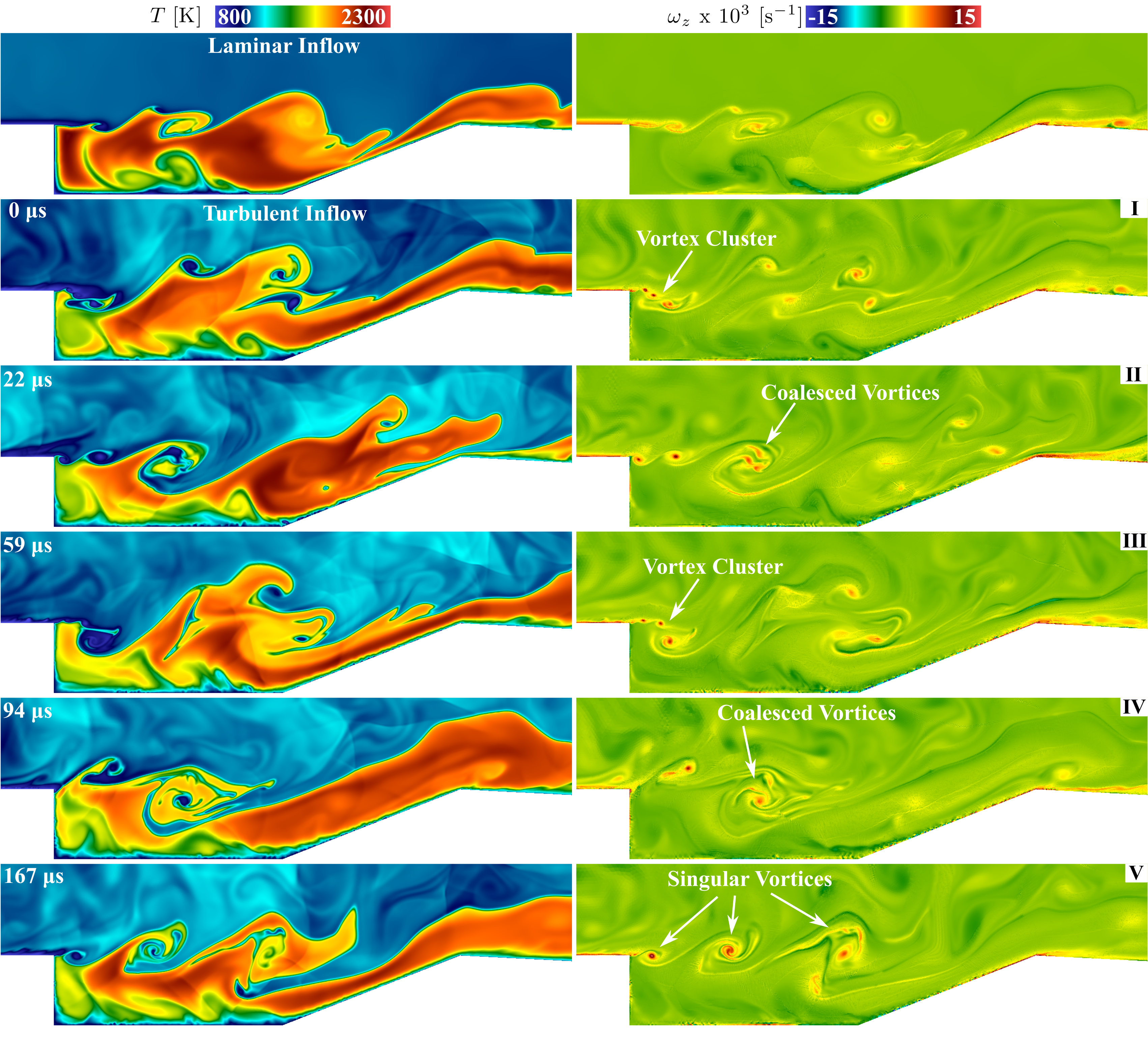}
	\caption{Temperature and $z$-vorticity contours for the laminar (top row) and turbulent inflow (remaining rows) cases. Several time steps are shown for the turbulent case.}
	\label{fig:vorticity}
\end{figure}

Figure~\ref{fig:vorticity} shows contours of temperature and $z$-vorticity, $\omega_\mathrm{z}$, where  $\omega_\mathrm{z}$ is a scalar field defined as $\frac{\partial v_y}{\partial x}  - \frac{\partial v_x}{\partial y}$. The figure provides a close-up view of the cavity for a single time step with the laminar inflow and several time steps with the turbulent inflow. In the laminar inflow plots, there are several distinct vortices visible in the cavity shear layer, reaching a peak vorticity of 10$^4$ s$^{-1}$ in the vortex centers. There is no vorticity visible above the cavity shear layer. 

The temperature and $\omega_\mathrm{z}$ contours are shown for five time steps in the turbulent inflow case. Elapsed time since the first image is shown in $\mu$s in the image corners, with the first image at 0 $\mu$s. In the image for 0 $\mu$s, there are several small vortices just downstream of the cavity leading edge that are propagating into the cavity shear layer, labeled as a ``Vortex Cluster" on the $\omega_\mathrm{z}$ image. As the vortex cluster moves further into the shear layer, the vortices orbit one another and coalesce into a single larger vortex. At 22 $\mu$s, the vortices have begun to coalesce, labeled ``Coalesced Vortices." At both 0 and 22 $\mu$s, a higher peak vorticity is observed than in the laminar inflow case, with peak vorticity in the vortex centers of 1.5 x 10$^4$ s$^{-1}$. Outside of the well defined vortices shown in the $\omega_\mathrm{z}$ contour plots for these two time steps, there is significant freestream vorticity above the cavity shear layer in the core flow that was not observed in the laminar inflow case also shown in Fig.~\ref{fig:vorticity}.

In the next two time steps, 59 and 94 $\mu$s, the same process occurs that was observed for the previous two time steps. Several small vortices shed from the cavity leading edge and coalesce into a single larger vortex that propagates through the cavity shear layer and downstream into the extender. This process repeated frequently throughout the course of the turbulent inflow simulation, with approximately 30 $\mu$s between shedding of the vortex cluster and coalescence into a single larger vortex. At 167 $\mu$s, several large ``Singular Vortices" are labeled. Each of these vortices is larger than the smaller vortices which coalesce. When the singular vortices are shed from the cavity leading edge, they are already the same approximate diameter as the vortices formed by several smaller vortices coalescing. The structures travel along the shear layer and gradually dissipate downstream in the extender. Their peak vorticity is the same as is observed for the smaller vortices, at approximately 1.5 x 10$^4$ s$^{-1}$. In the temperature contour plots in Fig.~\ref{fig:vorticity}, the vortices are also clearly seen as the flame is drawn into their recirculation regions. It is the convective vortex structures shedding from the cavity leading edge, and their subsequent breakup downstream in the cavity shear layer and extender, that are responsible for the significant wrinkling of the flame in this region. 

\begin{figure}
	\centering
	\includegraphics[width=86mm]{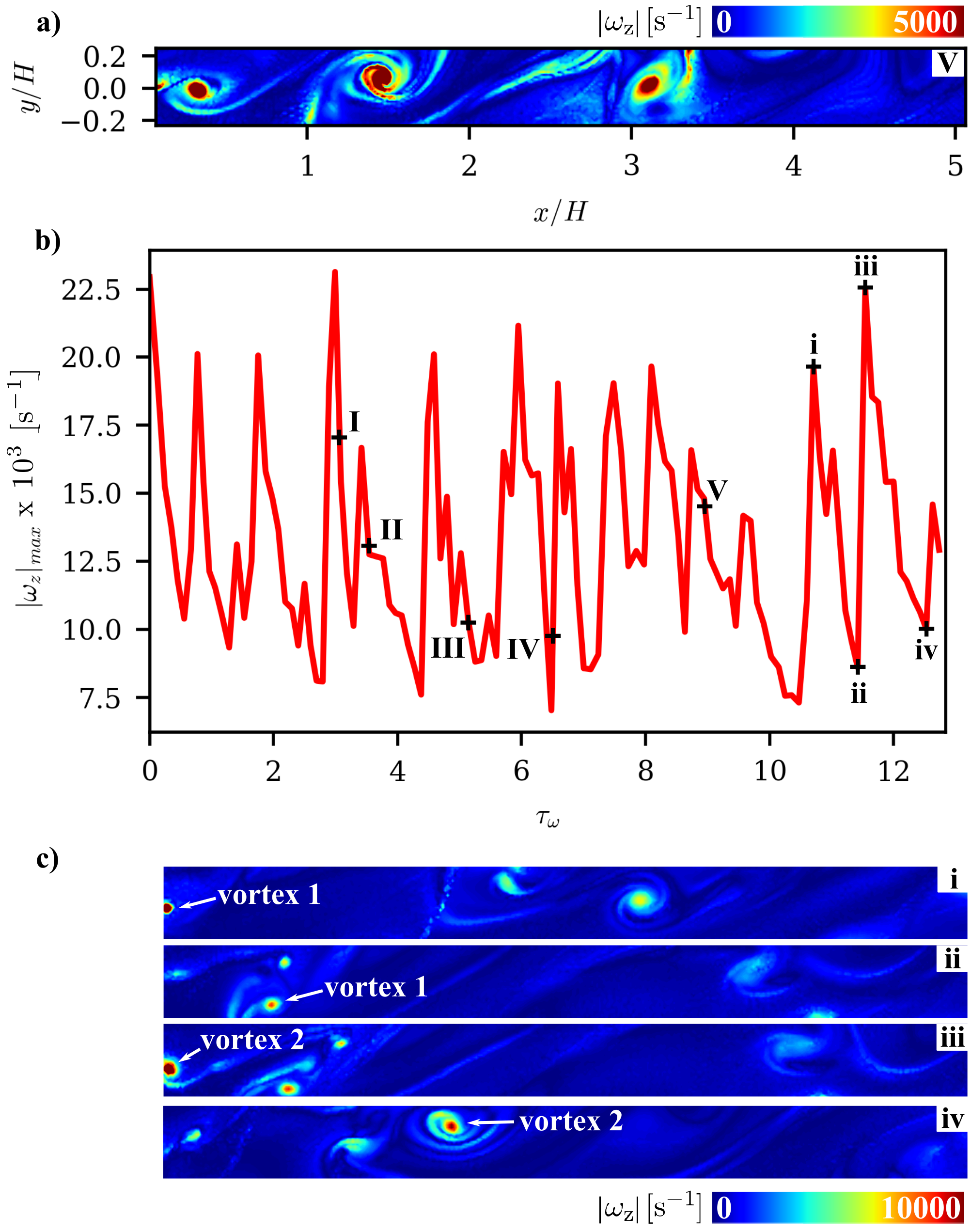}
	\caption{(a) $|\omega_\mathrm{z}|$ contours in the cavity shear layer for the time step labeled ``$\boldsymbol{\mathrm{V}}$" in Fig.~\ref{fig:vorticity} (b) Maximum $|\omega_\mathrm{z}|$ in the spatial domain shown in (a) during the turbulent inflow simulation and (c) $|\omega_\mathrm{z}|$ contours for four additional time steps to illustrate the vortex shedding process as related to maximum $|\omega_\mathrm{z}|$ generation.}
	\label{fig:shear_layer}
\end{figure}

Figure~\ref{fig:shear_layer} shows the maximum absolute value of $z$-vorticity, $|\omega_z|_{max}$, in the cavity shear layer, the region shown in Fig.~\ref{fig:shear_layer}(a), during the turbulent inflow simulation. The time step shown in Fig.~\ref{fig:shear_layer}a is the same as the last time step shown in Fig.~\ref{fig:vorticity}, marked with a ``$\boldsymbol{\mathrm{V}}$", where there are three vortices visible in the cavity shear layer. The $|\omega_z|_{max}$ values within this region are shown in Fig.~\ref{fig:shear_layer}(b) as a function of time normalized by the vortex shedding time scale, $\tau_\omega$, during the turbulent inflow simulation. The vortex shedding time scale is defined later in this section. It is evident that the $|\omega_z|_{max}$ is periodic with respect to time, varying between 7.5 and 22.5 x 10$^3$ s$^{-1}$. The time steps shown in Fig.~\ref{fig:vorticity} are marked ``$\boldsymbol{\mathrm{I}}$ - $\boldsymbol{\mathrm{V}}$" on Fig.~\ref{fig:shear_layer}(b). All of these time steps occur shortly after $|\omega_z|_{max}$ reaches a maximum, indicating that the maximum $z$-vorticity is observed when the vortices shown in the Fig.~\ref{fig:vorticity} images were shed from the cavity leading edge. 

\begin{figure}
	\centering
	\includegraphics[width=3.5in]{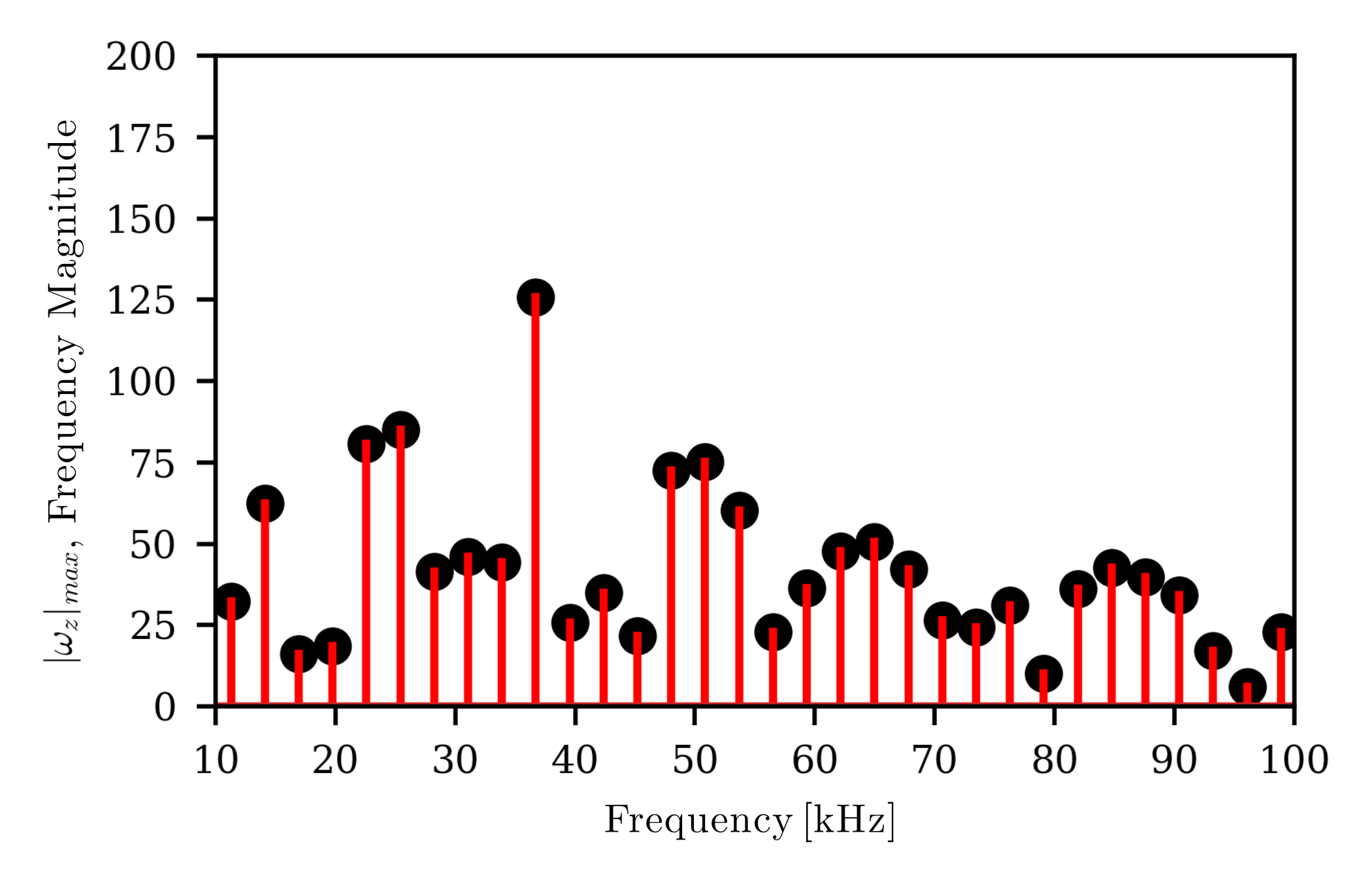}
	\caption{Fourier analysis of $|\omega_\mathrm{z}|_{max}$ showing the periodic data in the frequency domain.}
	\label{fig:vorticity_frequency}
\end{figure}

To explore the relationship between the vortex shedding process and the occurrence of maximum $z$-vorticity in the cavity shear layer, Fig.~\ref{fig:shear_layer}(c) shows contours of $|\omega_\mathrm{z}|$ for two local maxima and minima, labeled ``$\boldsymbol{\mathrm{i}}$ - $\boldsymbol{\mathrm{iv}}$" on Fig.~\ref{fig:shear_layer}(b). Frame $\boldsymbol{\mathrm{i}}$, a local $|\omega_\mathrm{z}|_{max}$ maximum, shows a vortex, labeled ``vortex 1" shedding from the cavity leading edge at $x/H$ = $y/H$ = 0. The vortex is relatively small in diameter, but has a concentrated region of high $|\omega_\mathrm{z}|$ at its center. There are several other vortices in the frame, all with lower peak vorticity than vortex 1, though larger in size. As the vortex leaves the cavity leading edge and propagates through the cavity shear layer, it grows from a small region of highly concentrated $|\omega_\mathrm{z}|$ to a larger region of moderate recirculation. In frame $\boldsymbol{\mathrm{ii}}$, a local $|\omega_\mathrm{z}|_{max}$ minimum, vortex 1 is shown, having propagated downstream of the cavity leading edge while peak vorticity dropped. A similar trend is shown for frames $\boldsymbol{\mathrm{iii}}$ and $\boldsymbol{\mathrm{iv}}$. A small region of high $|\omega_\mathrm{z}|$ sheds from the cavity leading edge, labeled ``vortex 2", resulting in a local $|\omega_\mathrm{z}|_{max}$ maximum. Several other smaller vortices are also visible. Vortex 2 travels into the shear layer where it grows in diameter as its peak vorticity decreases. This is shown in frame $\boldsymbol{\mathrm{iv}}$, corresponding to a local $|\omega_\mathrm{z}|_{max}$ minimum.

The metric $|\omega_\mathrm{z}|_{max}$ is therefore a reliable indicator for the vortex shedding cycle, where a local $|\omega_\mathrm{z}|_{max}$ maximum corresponds to a vortex being shed from the cavity leading edge. Thus, the periodicity of the $|\omega_\mathrm{z}|_{max}$ maximum is analogous to that of the periodic shedding of vortices. A Fourier analysis was performed to quantify the vortex shedding frequency. Figure~\ref{fig:vorticity_frequency} shows the discrete Fourier transform of the $|\omega_\mathrm{z}|_{max}$ data. The dominant frequency is clearly 37 kHz, correlating to the frequency at which a vortex sheds from the cavity leading edge. The inverse of this value provides the vortex shedding time scale of 27 $\mu$s, used to calculate $\tau_\omega$ in Fig.~\ref{fig:shear_layer}(b).

\subsubsection{Flame Strain Rate}

\begin{figure}[H]
	\begin{centering}
		\includegraphics[width=3in]{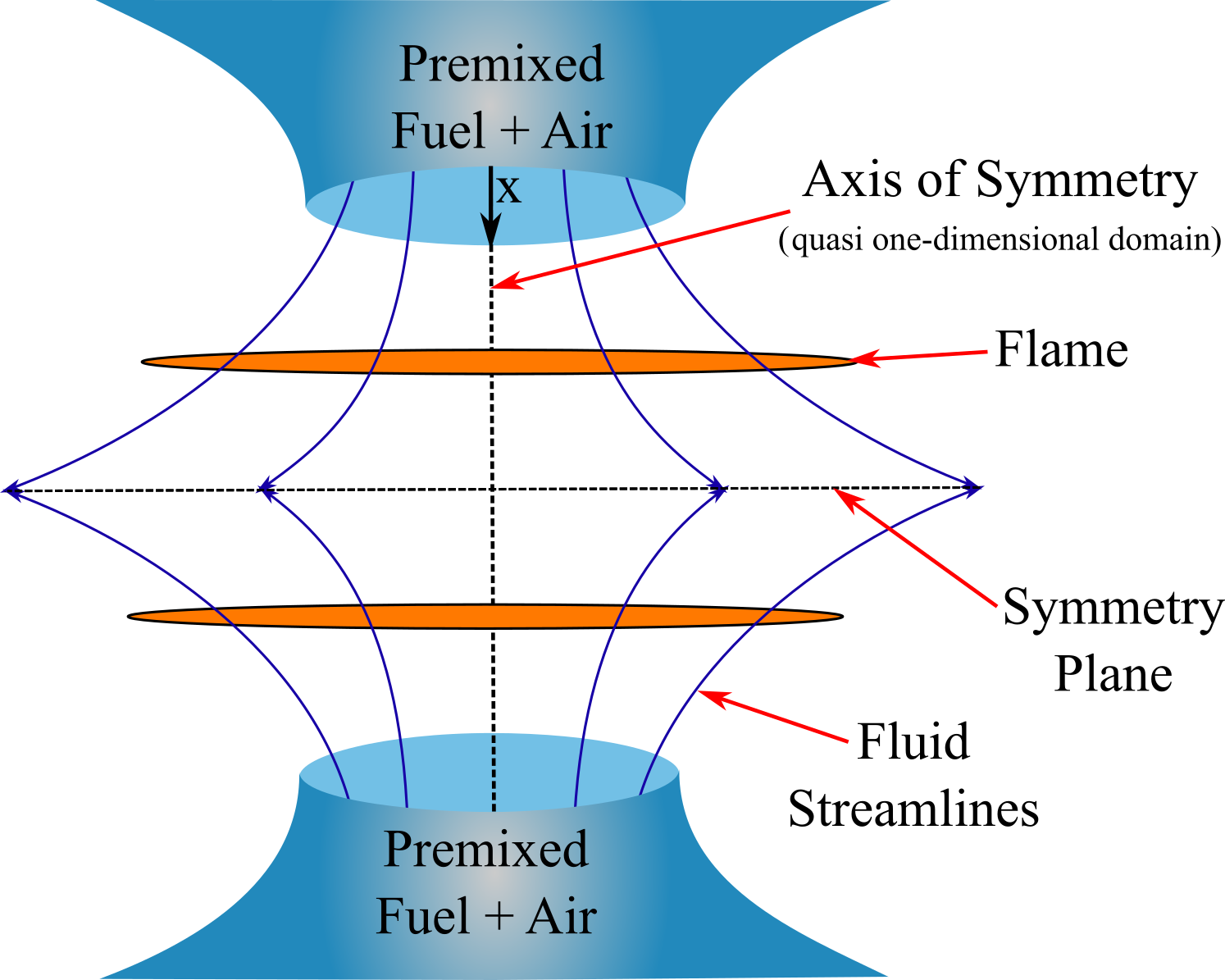}
		\par\end{centering}
	\caption{\label{fig:premixed_dual_flame_diagram} Premixed dual counterflow
		flame configuration: opposed nozzles with premixed air and fuel at
		the same mass flow rate create dual flames around a plane of symmetry
		that can be simulated using a steady quasi one-dimensional formulation.
		The formulation is derived along the axis of symmetry in the $x$-direction.}
\end{figure}
\begin{figure}[H]
	\begin{centering}
		\includegraphics[width=3.25in]{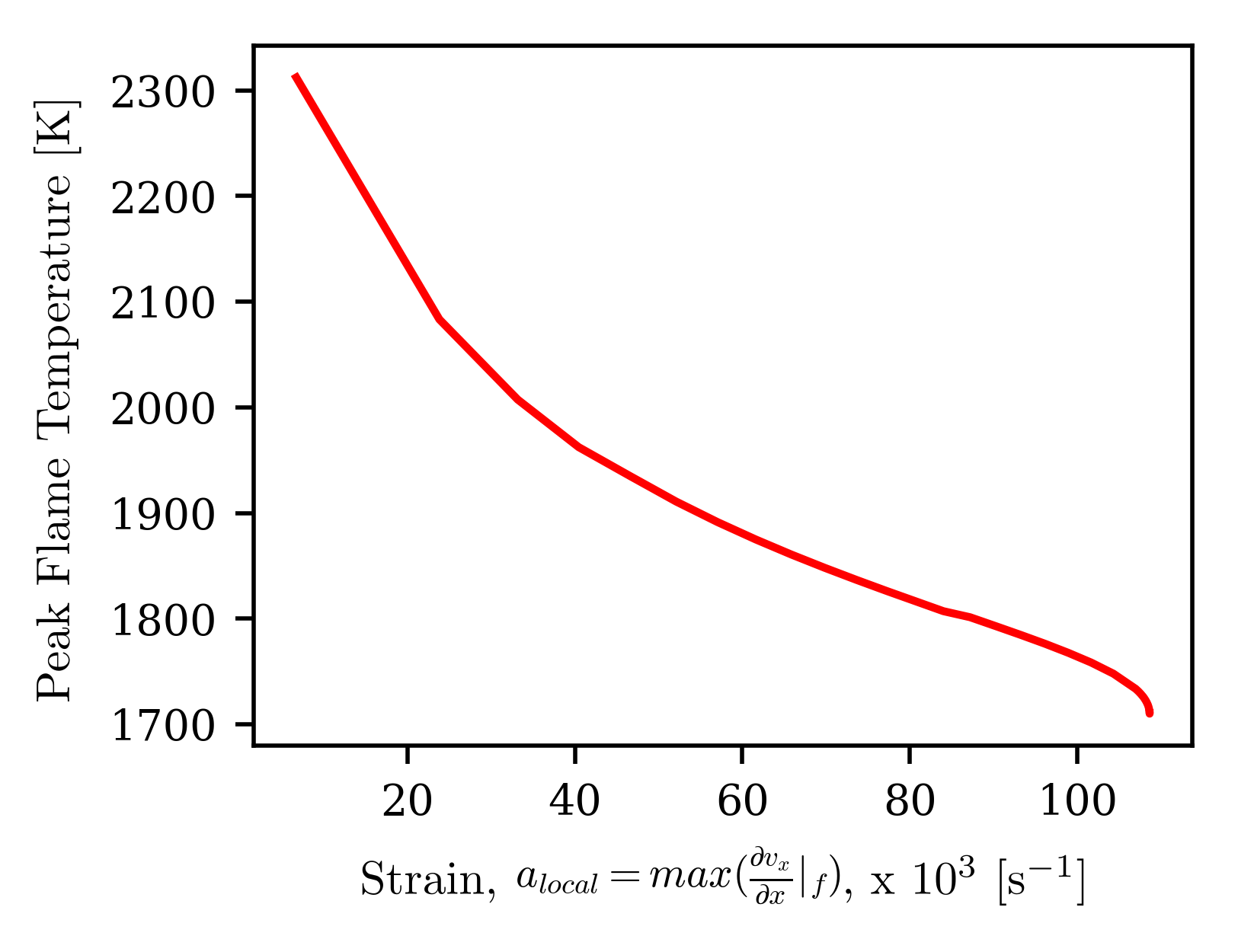}
		\par\end{centering}
	\caption{\label{fig:premixed_dual_flame_strain} Peak flame temperature as
		a function of local strain rate in premixed dual counterflow flame simulations
		were performed using the cavity inflow conditions. The extinction for this configuration occurs
		at 109 x 10$^{3}$ s$^{-1}$.}
\end{figure}

Premixed dual counterflow flame simulations were used to estimate
the extinction strain rate for the simulated cavity flameholder.
The premixed dual counterflow flame configuration is shown in Fig.~\ref{fig:premixed_dual_flame_diagram}. Two opposed jets containing
premixed fuel form a steady solution of two flames that are symmetric
about the stagnation plane. This configuration can be described by
a quasi one-dimensional formulation with the premixed fuel and air
and nozzle separation distance as free parameters. Using Cantera \cite{Cantera}
and the extinction strain curve approach outlined by \cite{Fia14},
several premixed dual counterflow flames of increasing strain were simulated. The cavity simulation inflow quantities were used as the
counterflow flame boundary condition, $\phi=0.6$, $T=1125$ K, and
$p=1.72$ atm along with an initial mass flux of 15 kg/s. Here
we define the strain rate as the local maximum velocity gradient prior
to the flame, $a_{local}=\max\left(\frac{\partial v_{x}}{\partial x}\vert_{f}\right)$,
which removes the geometric constraint on the counterflow solution
describing the extinction \cite{Che91}. Figure \ref{fig:premixed_dual_flame_strain}
shows the peak flame temperature as a function of strain rate for the cavity
conditions and reveals an extinction strain rate of 109 x 10$^{3}$ s$^{-1}$
for the chemical mechanism used in the cavity simulations \cite{dong2008numerical}.

\begin{figure}
	\centering
	\includegraphics[width=86mm]{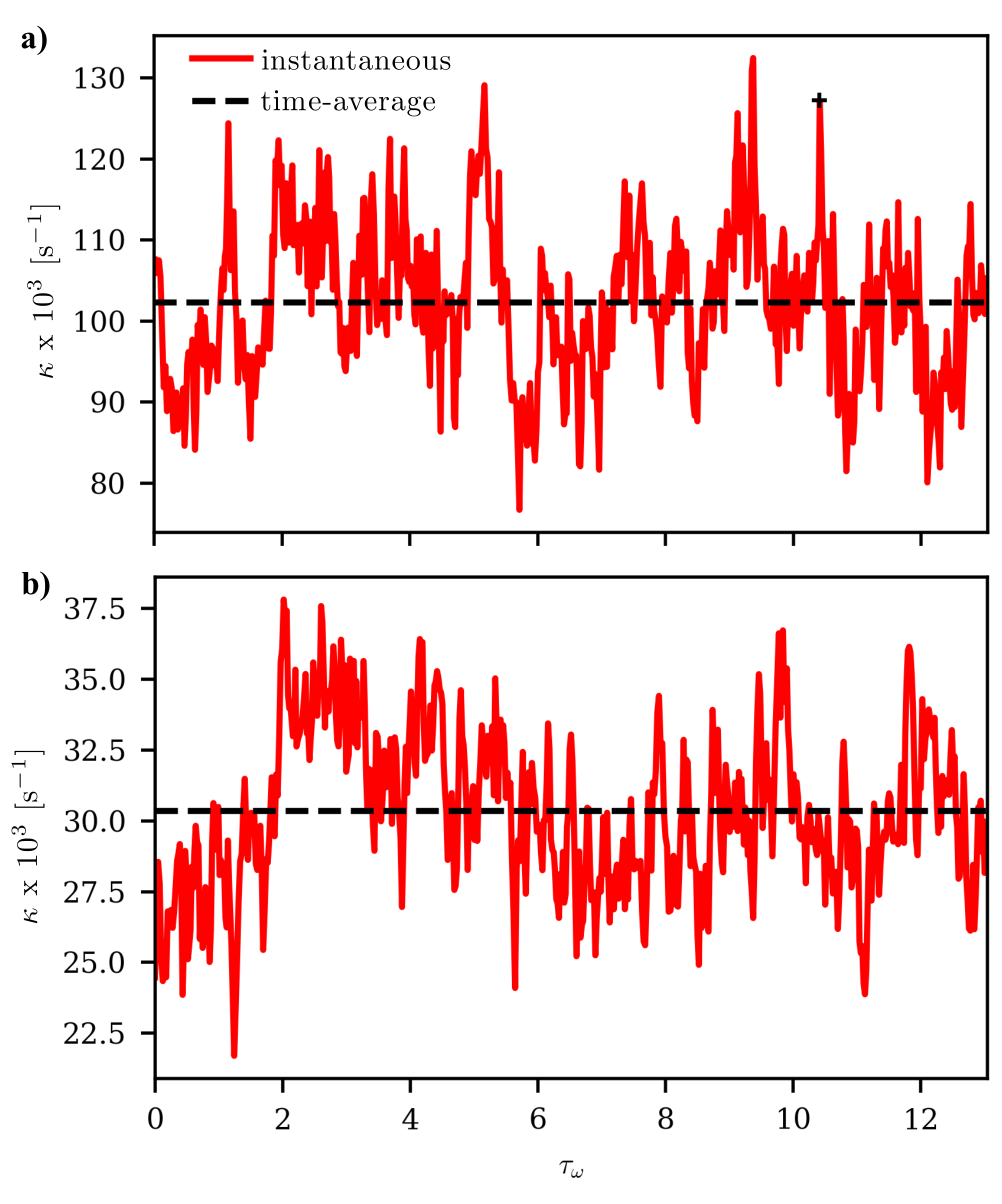}
	\caption{Mean strain rate along the flame surface as a function of time in the (a) cavity shear layer (0 $\leq$ $x/H$ $\leq$ 6.5) and (b) downstream of the cavity ($x/H$ $>$ 6.5). Total time-averaged strain rate across the plotted duration shown for each region. The ``$+$" at $\tau_\omega$ = 10.5 in (a) marks the time step shown in Fig.~\ref{fig:strain_contours}.}
	\label{fig:strain}
\end{figure}

Strain rate along the flame surface in the cavity flameholder, $\kappa$, was examined to understand where the flame is most highly strained and how the strain rate in the cavity combustor compares with the predicted extinction strain rate of 109 x 10$^{3}$ s$^{-1}$. Here, strain rate was calculated in accordance with the method of Poinsot and Candel \cite{poinsotBook2005,candel1990flame},

\begin{equation}
\kappa = (\delta_{ij} - n_i n_j)\frac{\partial w_i}{\partial x_j},
\end{equation}

\noindent where $\delta_{ij}$ is the Kronecker delta, ${n}$ is a component of the unit vector normal to the flame surface, $w$ is the flame speed, and $x$ is the spatial direction. Figure~\ref{fig:strain} shows mean strain rate as a function of time in the (a) cavity shear layer and in the (b) region downstream of the cavity. The time-averaged mean strain rate in each region is also shown. The flame is much more strained in the cavity than downstream, with time-averaged $\kappa$ three times greater in the cavity, where the time-averaged $\kappa$ of 102 x 10$^{3}$ s$^{-1}$ comes very close to matching the predicted extinction strain rate of 109 x 10$^{3}$ s$^{-1}$. This high strain rate leads to the noticeable stretching of the flame observed as it propagates through the cavity shear layer and into the extender, as discussed in Sec.~\ref{turb}. Downstream of the cavity, the mean strain rate decreases significantly below the predicted extinction value. In this region of lower strain rate, stretching of the flame is not observed to occur to the same degree that it does in the cavity shear layer. 

\begin{figure}
	\centering
	\includegraphics[width=165mm]{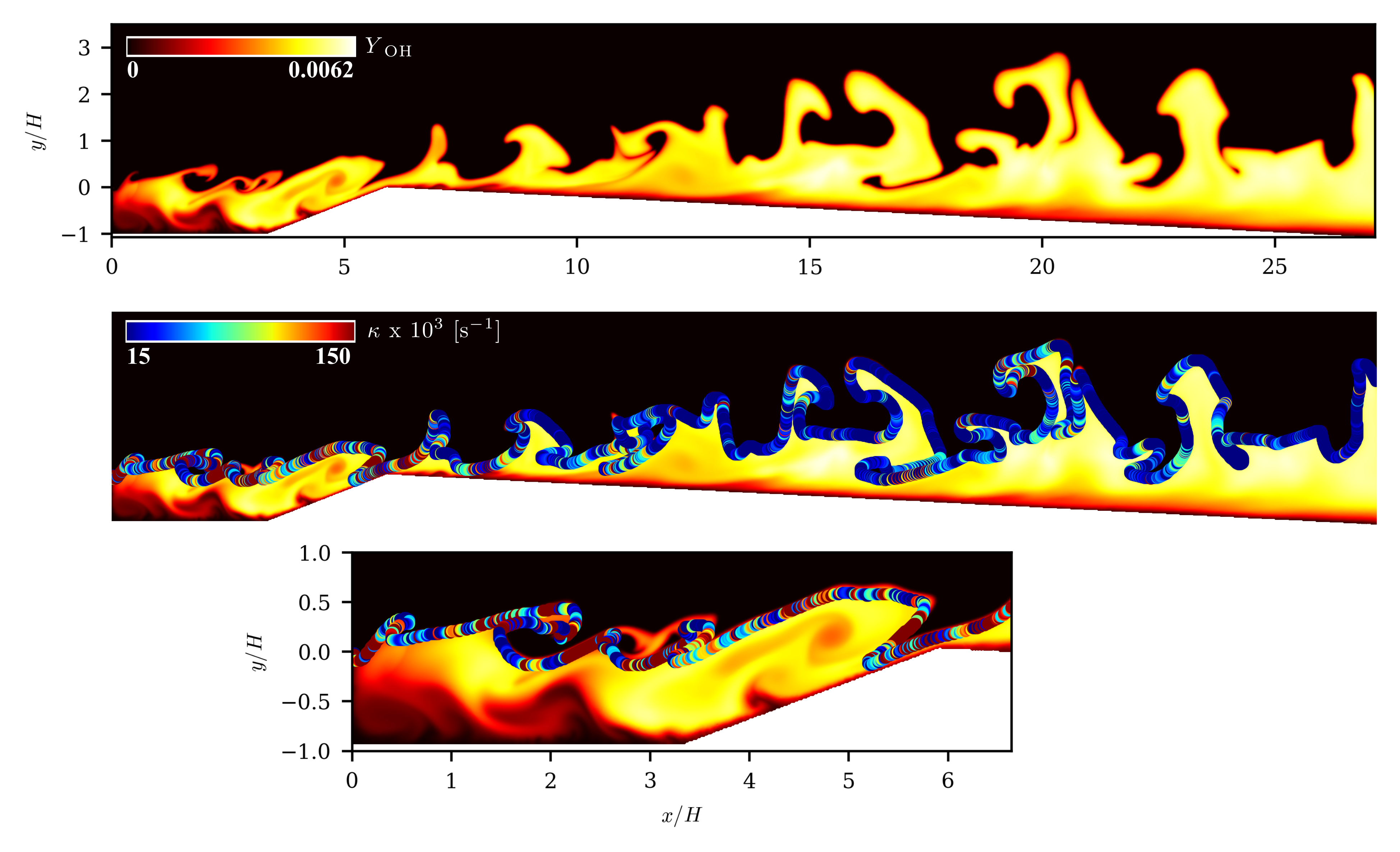}
	\caption{Contours of $Y_{\mathrm{OH}}$ overlaid with strain rate along the flame surface, with a close up view of the cavity.}
	\label{fig:strain_contours}
\end{figure}

Figure~\ref{fig:strain_contours} shows contours of $Y_{\mathrm{OH}}$ for a time step where mean strain rate in the cavity is at a local maximum. The time step shown in Fig.~\ref{fig:strain_contours} is marked on Fig.~\ref{fig:strain}(a) with a ``$+$", occurring at $\tau_\omega$ = 10.5. At this time step, there is considerable roll-up of the flame in the cavity shear layer, with several distinct vortices visible. Strain rate was calculated along the flame surface and overlaid on the  $Y_{\mathrm{OH}}$ contour plot in Fig.~\ref{fig:strain_contours}(b), with Fig.~\ref{fig:strain_contours}(c) providing a close-up view of the cavity region. Maximum $\kappa$ values are observed in the cavity shear layer, particularly just upstream of roll-up locations where the flame is being pulled into a vortex. Downstream of the cavity, $\kappa$ is significantly diminished as the flame propagates out of the cavity shear layer and into the core flow. However, local regions of relatively increased $\kappa$ are still visible in the extender where the flame is highly curved and roll-up occurs. 

\begin{figure}
	\centering
	\includegraphics[width=85mm]{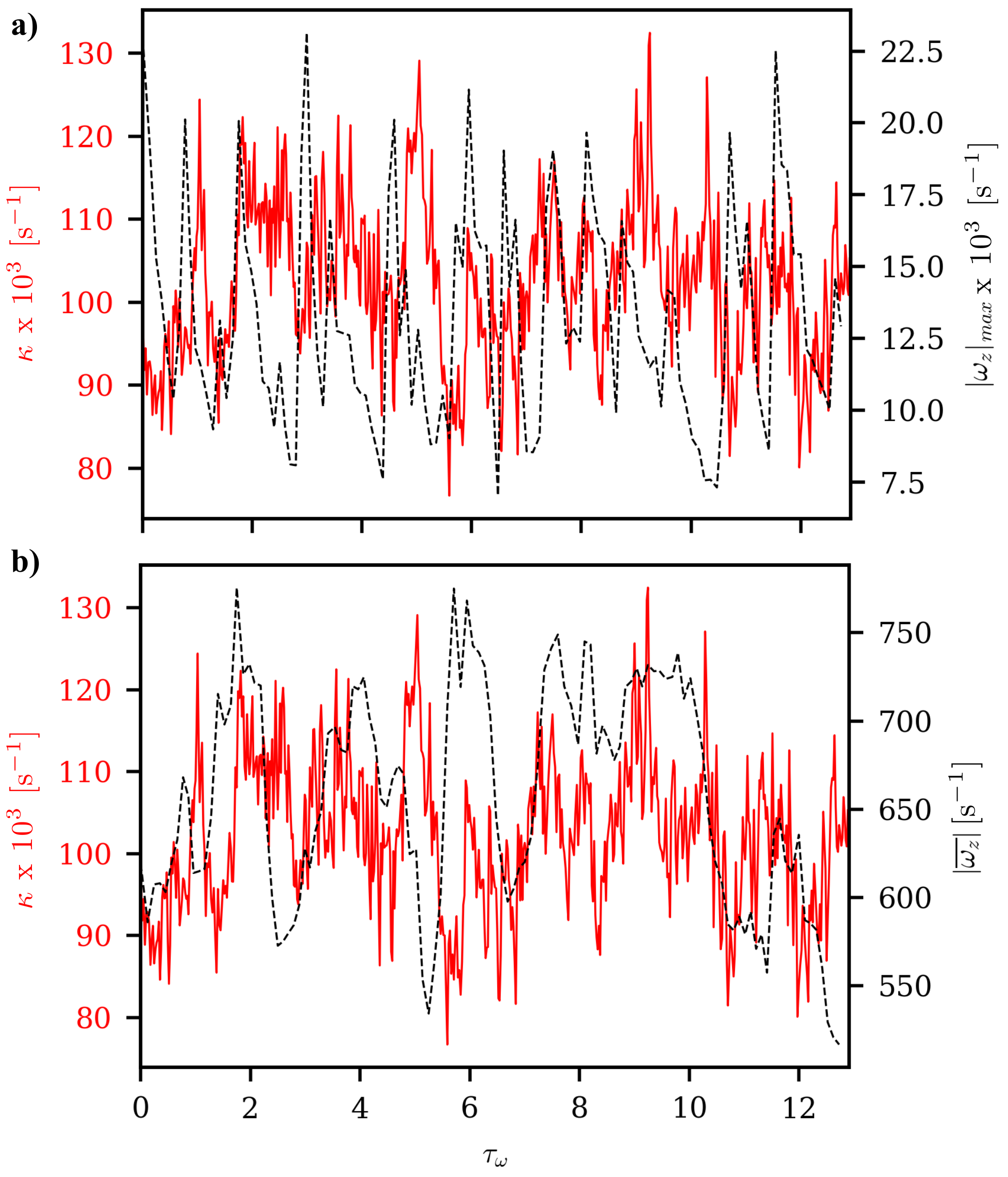}
	\caption{Mean strain rate along the flame surface in the cavity as a function of time, overlaid with (a) the absolute value of maximum vorticity in the cavity shear layer and (b) the absolute value of average vorticity in the cavity shear layer.}
	\label{fig:strain_vorticity}
\end{figure}

\begin{figure}
	\centering
	\includegraphics[width=3.5in]{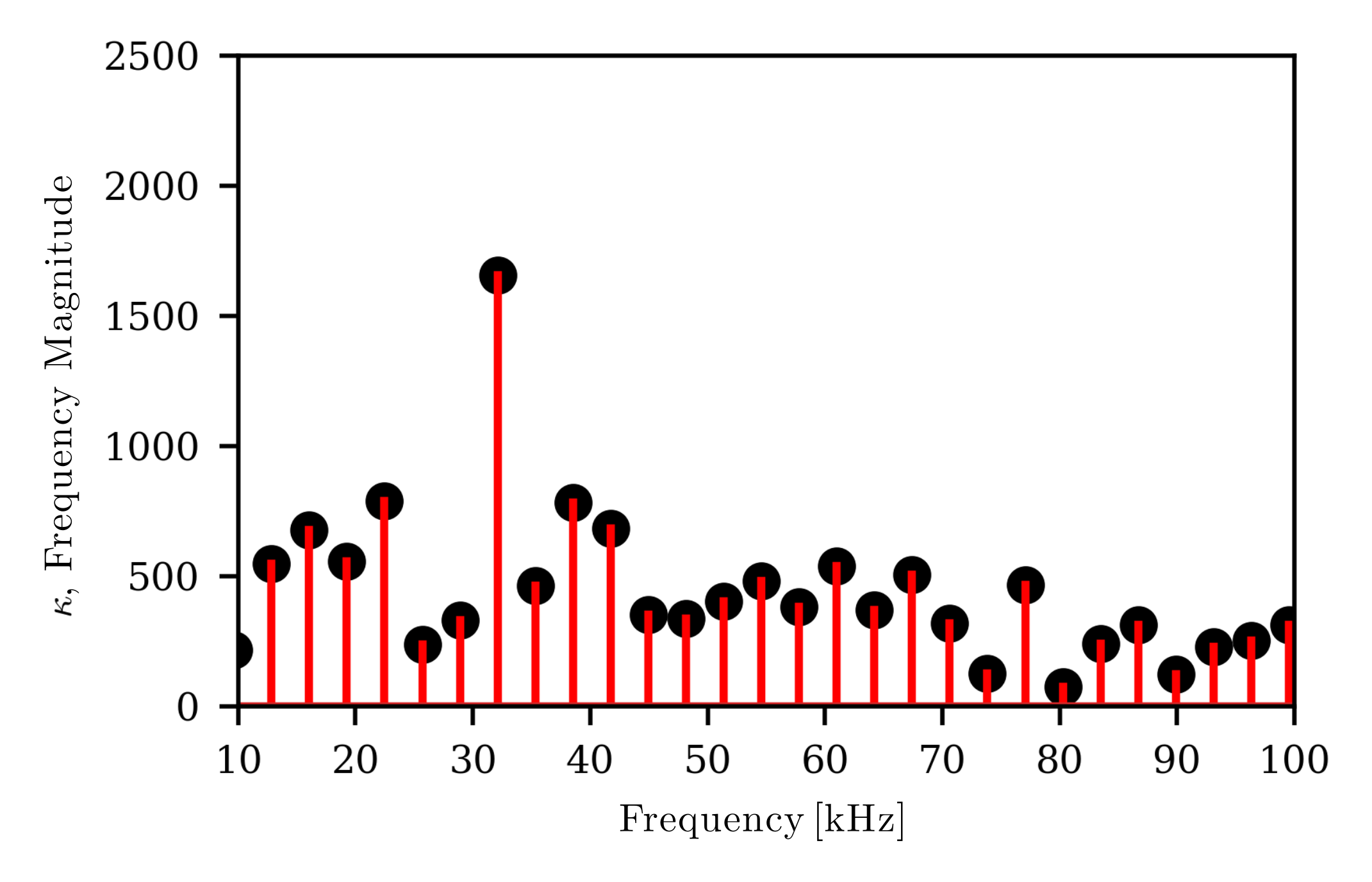}
	\caption{Fourier analysis of mean $\kappa$ in the cavity (from Fig.~\ref{fig:strain}(a)) showing the periodic data in the frequency domain.}
	\label{fig:strain_frequency}
\end{figure}

As was the case for $|\omega_\mathrm{z}|_{max}$, the mean $\kappa$ in the cavity shear layer, shown for the duration of the turbulent inflow simulation in Fig.~\ref{fig:strain}(a), appears to be periodic with local maxima occurring at a regular frequency. Figure~\ref{fig:strain_vorticity}(a) shows the mean cavity $\kappa$ as a function of time, overlaid with $|\omega_\mathrm{z}|_{max}$. Most of the peaks in $\kappa$ are preceded by a peak in $|\omega_\mathrm{z}|_{max}$ with a delay between the peaks of several microseconds. Thus it is apparent that the vortices shed from the cavity leading edge, as predicted by $|\omega_\mathrm{z}|_{max}$, drive the local $\kappa$ in the cavity. This is a logical result, considering that high mean $\kappa$ results from localized shear as the flame rolls up, is stretched, and pulled into the vortices. Figure~\ref{fig:strain_vorticity}(b) shows the same mean $\kappa$ as in (a), here overlaid with mean $z$-vorticity, $|\overline{\omega_z}|$, in the cavity shear layer. To obtain $|\overline{\omega_z}|$, the $z$-vorticity was spatially averaged across the same domain as that used to obtain $|\omega_\mathrm{z}|_{max}$, shown in Fig.~\ref{fig:shear_layer}(a). Unlike $|\omega_\mathrm{z}|_{max}$, the $|\overline{\omega_z}|$ data does not appear to be periodic, but it does appear to track the general trends of increasing and decreasing $\kappa$.

A Fourier analysis was performed to determine the dominant frequency in the mean cavity $\kappa$ data. Figure~\ref{fig:strain_frequency} shows mean cavity $\kappa$ in the frequency domain. The dominant frequency is 32 kHz, which is quite close to the dominant frequency in $|\omega_\mathrm{z}|_{max}$ of 37 kHz, confirming the observation that local strain rate in the flame in the cavity shear layer is driven by the cavity's vortex shedding process.

\subsubsection{Pressure Fluctuations} \label{pressure_fluct}

\begin{figure}
	\centering
	\includegraphics[width=144mm]{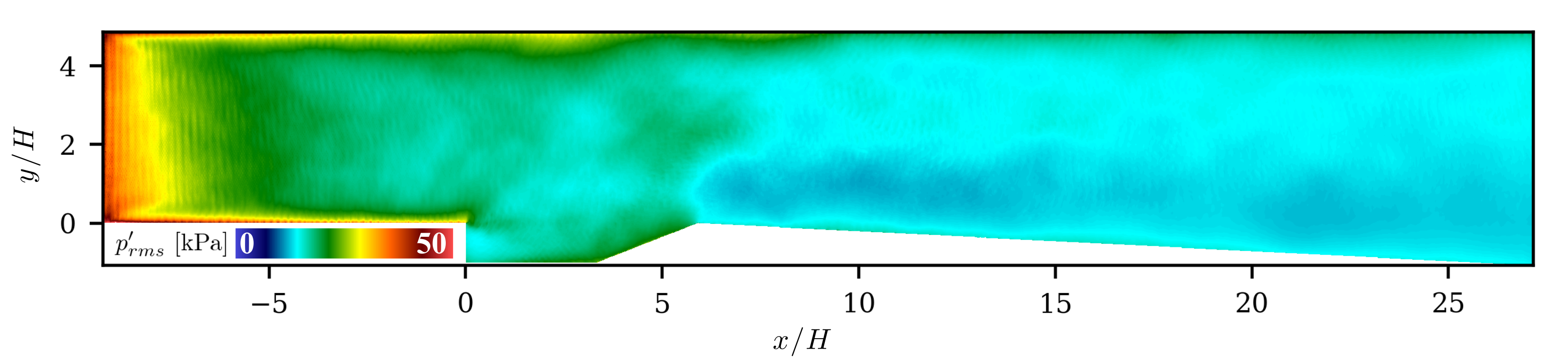}
	\caption{Contours of $p^\prime_{rms}$ for the 2D turbulent inflow simulation.}
	\label{fig:pressure_fluc_contour}
\end{figure}

In order to examine any relationship between flame strain rate and pressure fluctuations in the cavity, the static pressure fluctuations from the mean, $p^\prime$, were tracked for duration of the turbulent inflow simulation. Figure~\ref{fig:pressure_fluc_contour} shows the root-mean-square of $p^\prime$ in the entire domain. The trend is very similar to that of $TI$ shown in Fig.~\ref{fig:turbulent_flame}, where the fluctuations initially decay downstream of the inflow to a relatively steady value upstream of the cavity before decaying again downstream of the cavity through the extender. The magnitude of pressure fluctuations upstream and within the cavity is approximately 25 kPa, which is 15\% of the inflow static pressure. 

\begin{figure}
	\centering
	\includegraphics[width=88mm]{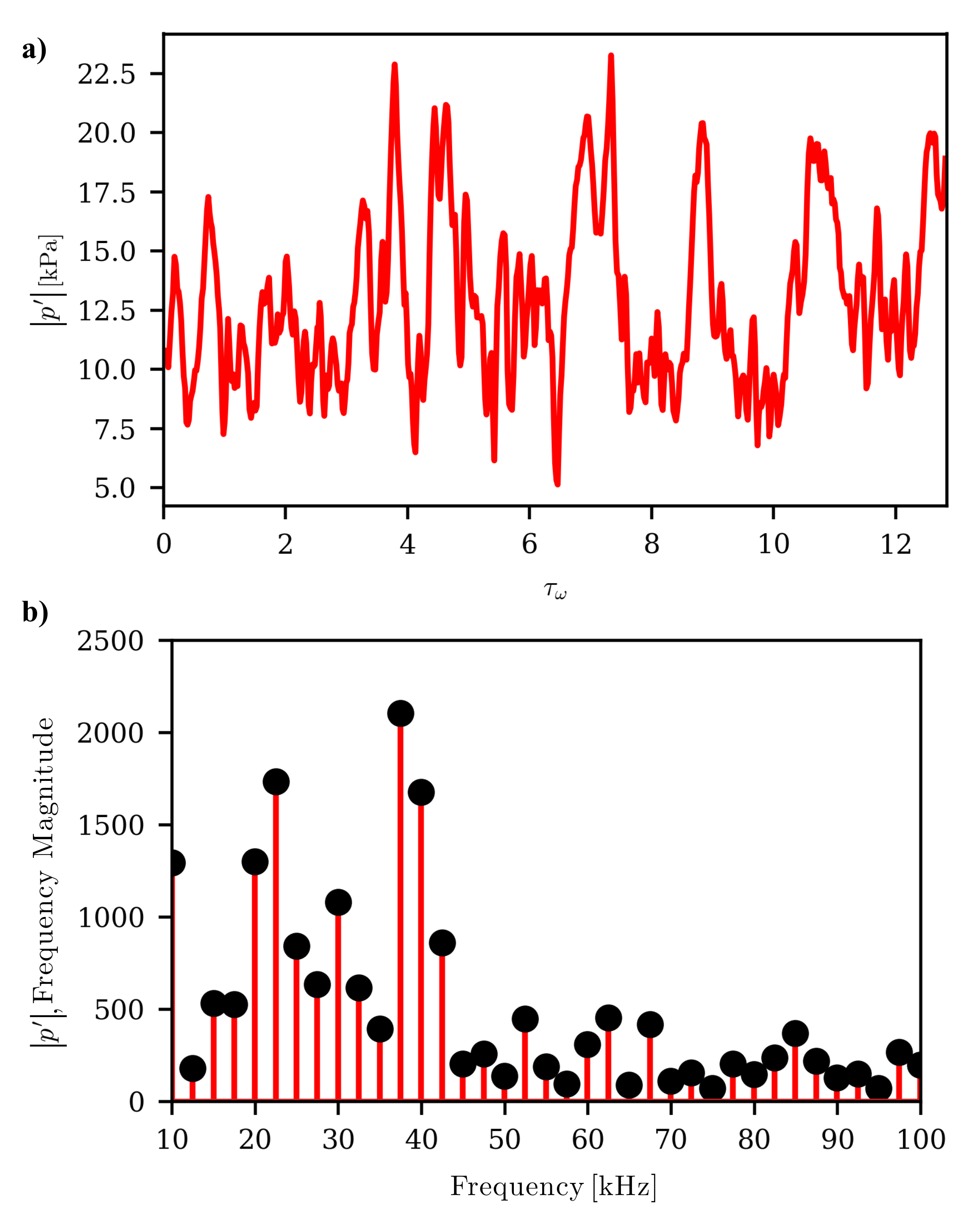}
	\caption{(a) Mean pressure fluctuation in the cavity shear layer as a function of time and (b) in the frequency domain.}
	\label{fig:pressure_modal_analysis}
\end{figure}

Figure~\ref{fig:pressure_modal_analysis}(a) shows the absolute value of mean pressure fluctuation in the cavity shear layer during the turbulent inflow simulation. The spatial domain over which this measurement was taken is the same as that used in the measurement of vorticity in the cavity shear layer, shown in Fig.~\ref{fig:shear_layer}(a). The pressure fluctuates periodically, with mean $|p^\prime|$ ranging from 5 to 22.5 kPa in the cavity shear layer. A Fourier analysis was performed to determine the dominant frequency in the $|p^\prime|$ data. Figure~\ref{fig:pressure_modal_analysis}(b) shows $|p^\prime|$ in the frequency domain. The dominant frequency is 37 kHz, the same as that of $|\omega_\mathrm{z}|_{max}$ and very close to the dominant frequency for mean $\kappa$ in the cavity shear layer of 32 kHz. This relationship illustrates that not only is the strain rate of the flame driven by the cavity's vortex shedding process, but so too are the pressure fluctuations in the cavity shear layer. Thus $|p^\prime|$ is also a reliable indicator of vortex shedding frequency and periodicity in $\kappa$.

\subsubsection{Higher Order Solution} \label{high_order}

\begin{figure}
	\centering
	\includegraphics[width=144mm]{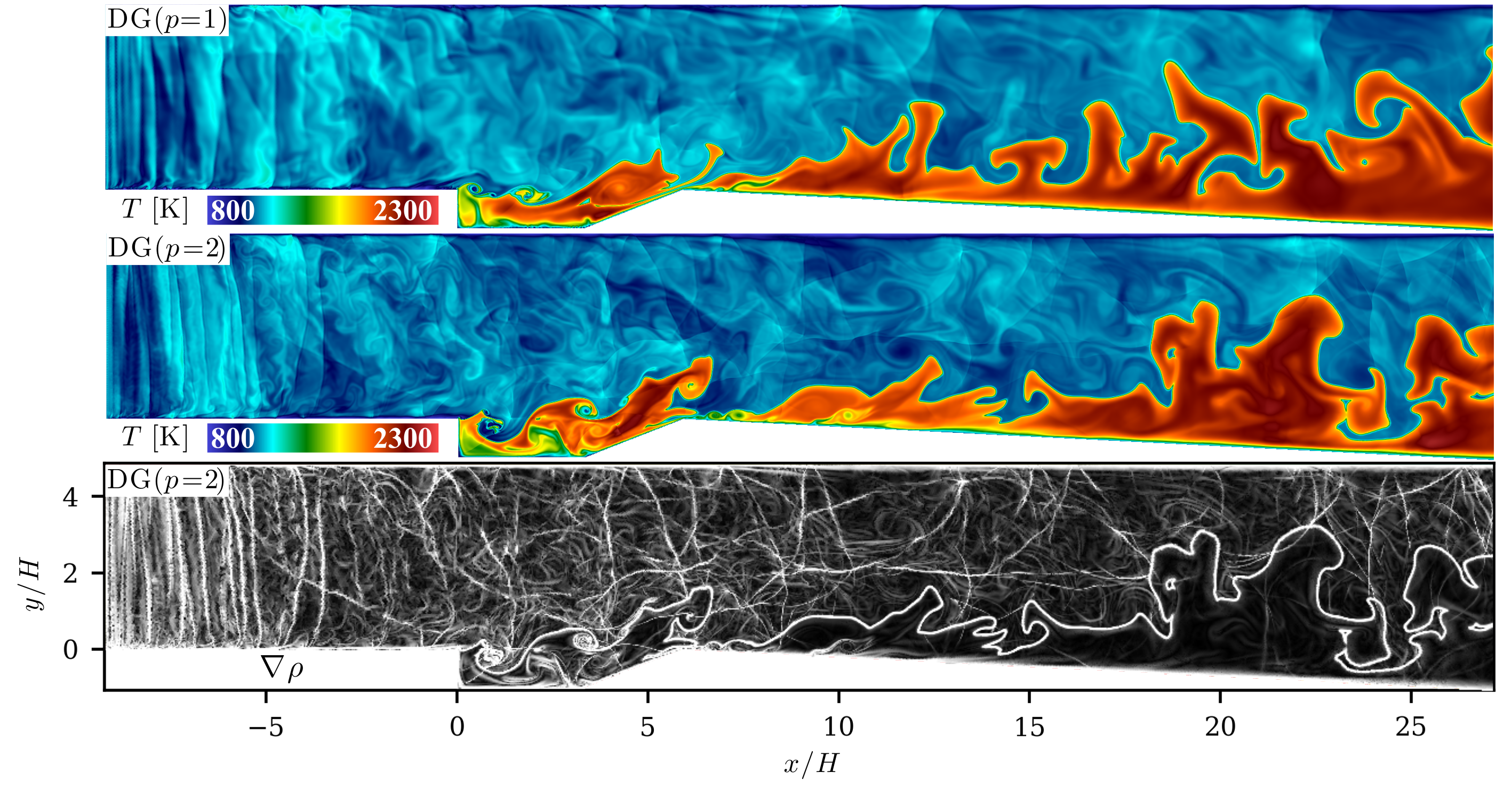}
	\caption{Temperature contours for DG($p=1$) and DG($p=2$) solutions and numerical schlieren contours for DG($p=2$) solution.}
	\label{fig:p2_solution}
\end{figure}

A subsequent simulation was performed using the same mesh element sizes as used in the previously described cases, but with third-order accurate DG($p=2$) elements. In this configuration, the mesh resolution is approximately 10 $\mu$m in the cavity shear layer, flame, and against the walls, growing to 75 $\mu$m in the core flow. This simulation was performed in order to assess differences in the level of detail captured in $\mathrm{DG}(p=1)$ and $\mathrm{DG}(p=2)$ solutions. The final time step of the DG($p=1$) solution was used as the initial condition for the DG($p=2$) simulation. Figure~\ref{fig:p2_solution} shows temperature and numerical schlieren contours for the DG($p=2$) solution, with the temperature contours for the DG($p=1$) solution shown above for comparison. In comparing the temperature contours for the DG($p=1$) and DG($p=2$) solutions, it is apparent the the DG($p=2$) solution resolves the acoustic waves generated by the turbulent inflow with more detail. However, the two simulations show similar structure along the flame surface and both resolve the local stretching of the flame as it propagates over the cavity ramp. In both cases, the flame remains anchored in the cavity and, downstream of the cavity, travels a comparable distance from the bottom wall into the core flow resulting in a consistent flame angle. The DG($p=1$) solution is thus considered adequate for capturing the turbulence-flame interactions and resolving the physics of interest in this study.

\subsection{Three-Dimensional Simulations} \label{3D}

\begin{figure}
	\centering
	\includegraphics[width=165mm]{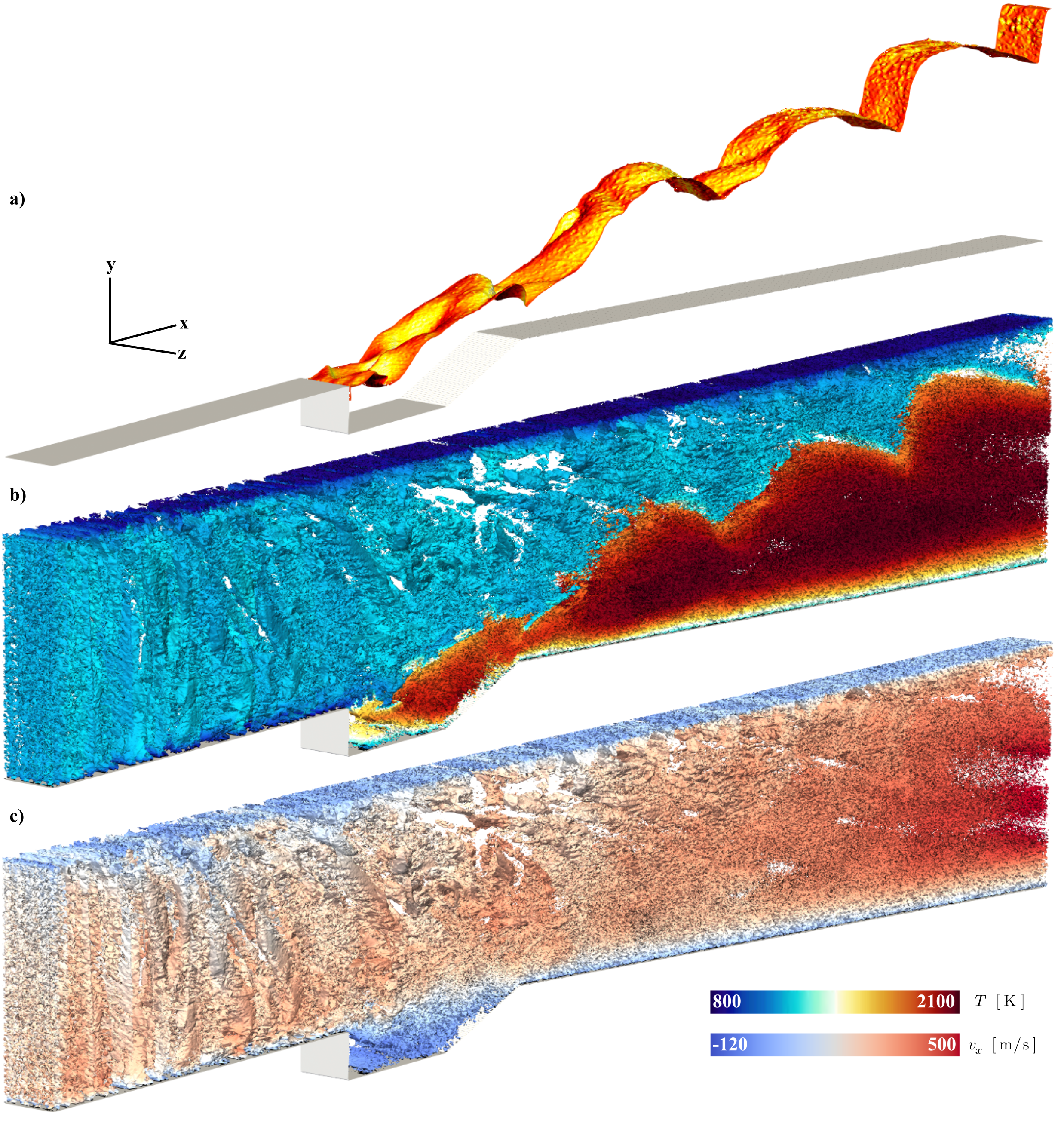}
	\caption{Results from the 3D simulation showing (a) an isosurface of $T$ = 1300 K (b) Q-criterion isosurfaces colored by $T$ and (c) Q-criterion isosurfaces colored by $v_x$. }
	\label{fig:3D}
\end{figure}

A 3D simulation was performed using the turbulent inflow boundary condition. The mesh was designed to give maximum possible resolution while still being computationally affordable on the desktop GPU machine on which the calculations were performed. The $x$ and $y$ dimensions are the same as those in the 2D simulation and a 4 mm $z$-span is used with periodic boundaries. The upper and lower walls are isothermal, with the same temperatures used in the 2D simulations. The mesh element sizes are 140 $\mu$m in the cavity shear layer and flame, 200 $\mu$m in the boundary layers, and 440 $\mu$m in the core flow. DG($p=2$) elements were used, providing approximate resolutions of 45 $\mu$m in the cavity shear layer and flame, 65 $\mu$m in the boundary layers, and 145 $\mu$m in the core flow. As described in Sec.~\ref{turb_inflow}, the turbulence generated at the inflow boundary is isotropic, with the same magnitude of velocity perturbation applied in all three directions. No mean $z$-velocity profile was available from the experiments or prior simulations. Instead, the initial $z$-velocity at the inflow plane was prescribed with a sine wave, 

\begin{equation}
v_z (t) = A\sin({P\frac{2\pi}{w} z}),
\end{equation}

\noindent where $A$ is the wave amplitude, $P$ is the period, and $w$ is the $z$-span width. For this simulation, $A$ and $P$ were chosen to be 100 m/s and 4, respectively. Additional simulations were performed with lower values for wave amplitude and fewer periods; in these cases there was less variation in the flame surface in the $z$-direction and so the case with $A$ = 100 m/s and $P$ = 4 was selected as the most interesting for discussion here. 

Figure~\ref{fig:3D} shows results from the 3D simulation. Figure~\ref{fig:3D}(a) is an isosurface of 1300 K, used to show the flame surface. It is colored by magnitude of the $z$-component of the surface normal vector to show surface texture. The flame is highly wrinkled, particularly in the cavity shear layer, gradually becoming more smooth downstream of the cavity with good penetration into the core flow. In the cavity shear layer the flame is highly three-dimensional, with considerable variation in the $z$-direction. Figures~\ref{fig:3D}(b) and (c) show isosurfaces of $Q$-criterion colored by temperature and $x$-velocity, respectively. The $Q$-criterion is commonly used to identify and visualize vortices, defined as connected fluid with a positive second invariant of $\nabla \boldsymbol{v}$ and where the vorticity magnitude is greater than the strain rate magnitude,

\begin{equation}
Q \equiv \frac{1}{2}(v^2_{i,i}-v_{i,j}v_{j,i}) > 0,
\end{equation}

\noindent where the comma in the subscript denotes differentiation \cite{kolavr2007vortex}. Figures~\ref{fig:3D}(c) and (b) show isosurfaces of $Q$ = 1e7, selected to visualize an abundance of vortices of varying length scales throughout the domain. In Figs.~\ref{fig:3D}(b) and (c), it is evident that the vortices generated by the turbulent inflow propagate into the domain at a relatively constant size in the core flow, with variability in size increasing as the flow gets closer to the cavity. The average vortex size also decreases with distance to the upper and lower walls. The most variability in vortex size is observed above the cavity, where the vortices are large in the core flow relative to that observed in the cavity and near the top wall. As the vortices pass through the flame front, the size decreases notably and approaches uniformity toward the outflow boundary where the flow primarily consists of products of combustion. This indicates that the turbulence dissipates to smaller scales as the fuel is processed by the flame.

\begin{figure}
	\centering
	\includegraphics[width=70mm]{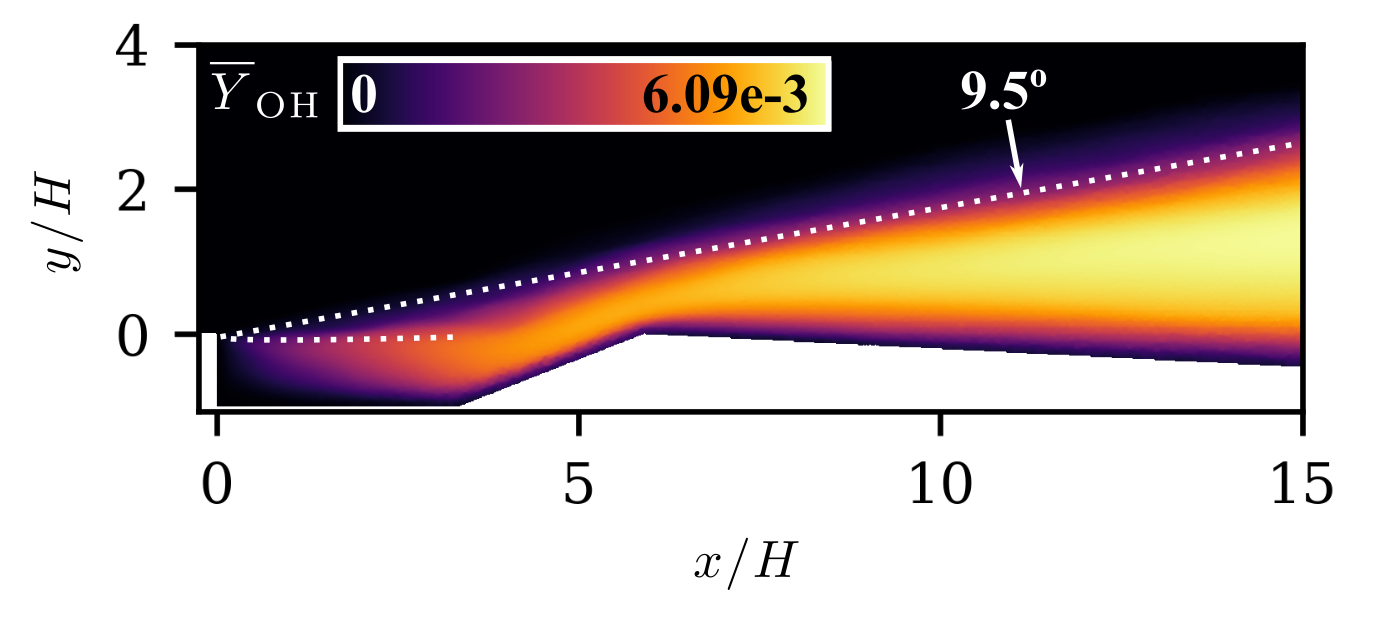}
	\caption{$\overline{Y}_\mathrm{OH}$ from 3D simulation with flame angle shown.}
	\label{fig:OH_3D}
\end{figure}

The flame remained anchored to the cavity leading edge for the duration of the simulation with significant lift off the bottom wall of the extender. Figure~\ref{fig:OH_3D} shows $\overline{Y}_\mathrm{OH}$ for the 3D simulation, taken at the center plane of the $z$-span. The flame angle agrees with the theoretical flame angle, shown in Fig.~\ref{fig:oh_comparison}(a), and is slightly larger than that of the 2D simulation with turbulent inflow. This is likely due in part to the somewhat decreased spatial resolution in the 3D simulation, compared to that of the 2D case, which results in increased thermal diffusion from the flame front into the inflowing reactants. 

\begin{figure}
	\centering
	\includegraphics[width=145mm]{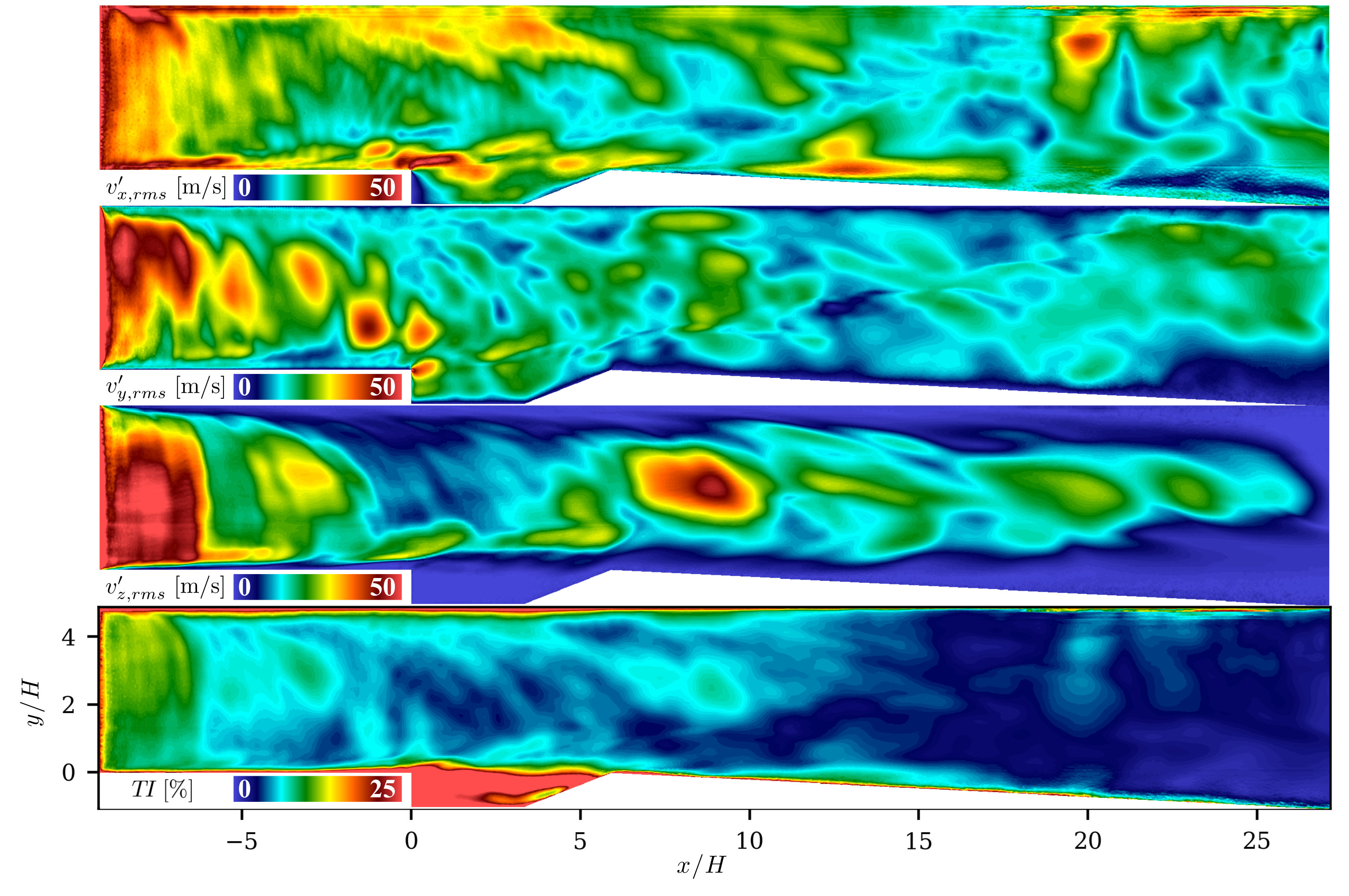}
	\caption{Contours of the root-mean-square of velocity perturbations and turbulence intensity at the mid-plane of the $z$ span in the 2 mm $z$-span width 3D case.}
	\label{fig:3D_turbulent_flame}
\end{figure}

A second 3D simulation was performed using a domain with a 2 mm $z$-span width. This reduced volume compared to the previously discussed 3D case allowed for increased spatial resolution. This calculation used a DG($p=2$) mesh with element sizes of 90 $\mu$m in the cavity shear layer and flame, 130 $\mu$m in the boundary layers, and 290 $\mu$m in the core flow, corresponding to approximate resolutions of 30, 45, and 95 $\mu$m, respectively. Figure~\ref{fig:3D_turbulent_flame} shows contour plots of $v'_\mathrm{x,rms}$, $v'_\mathrm{y,rms}$, $v'_\mathrm{z,rms}$, and $TI$ in the middle of the $z$-span at $z$ = 1 mm. As was observed in the 2D case, high-frequency waves generated by the turbulent inflow propagate into the channel and are observable features in the $v'_\mathrm{x,rms}$ plot. The magnitude of $v'_\mathrm{x,rms}$ is greatest near the walls and in the cavity shear layer, with a thin region of particularly high $v'_\mathrm{x,rms}$ beginning at the cavity leading edge and propagating into the cavity shear layer. In all three directions, the velocity perturbations range in magnitude from 10 to 50 m/s upstream of the cavity, decaying slightly downstream of the cavity in the extender. Turbulence intensity ranges from 10 to 15\% upstream of the cavity, also decreasing downstream. The highest levels of $TI$ are observed near the walls and in the cavity. These features are in agreement with the 2D turbulent inflow case.


\section{Conclusions} \label{conclusions}

This paper summarizes the numerical simulation of turbulent combustion in the University of Virginia Supersonic Combustion Facility using a synthetic turbulence inflow boundary condition to accurately and efficiently reproduce the turbulence intensity measured in experiments. Ramjet-mode operation of the facility at a combustor inflow Mach number of 0.6 was simulated, with a stagnation temperature of 1200 K corresponding to flight at Mach 5. The inflow was premixed ethylene-air with an equivalence ratio of 0.6 and a 19 species, 35 reaction step chemical mechanism was used to simulate ethylene-air combustion. Both 2D and 3D simulations were performed. Two 2D cases were considered: (a) laminar inflow and (b) synthetic turbulence inflow. In both cases, a cavity-stabilized flame was achieved. The turbulent inflow boundary condition successfully reproduced the experimental measurements of the turbulent flow profile upstream of the combustor cavity. This allowed the accurate simulation of the fine-scale combustion in the cavity flameholder without incurring the computational expense of simulating the entire upstream domain. 

Due to the high resolution in the 2D simulations, local stretching of the flame as it propagated over the cavity ramp was observed, supporting the experimental observations. In the laminar inflow case, the flame angle was much shallower than that measured experimentally. The flame angle in the turbulent inflow case agreed with the experiment and with theory, indicating that the inflow turbulence has a significant effect on flame propagation into the core flow. The inflow turbulence promoted robust combustion as more of the incoming fuel was consumed in a shorter axial span than in the laminar inflow case. 

Vorticity was measured in the cavity shear layer during the turbulent inflow simulation and maximum vorticity was observed to occur periodically, coinciding with a vortex shedding event. A discrete Fourier transform of the maximum vorticity data showed a dominant frequency of 37 kHz, correlating to the vortex shedding frequency during the turbulent inflow simulation. The strain rate along the flame surface was also calculated and tracked, showing much higher strain rate in the portion of the flame in the cavity shear layer than downstream in the extender. The strain rate in the cavity shear layer tracks the predicted extinction strain rate for the combustor's inflow conditions. Highest strain rate was observed just upstream of the flame roll-up locations in the cavity shear layer, contributing to the local stretching observed as the flame propagates over the cavity ramp. A Fourier analysis of the periodic mean strain rate data revealed that local maxima occur at the vortex shedding frequency, indicating that the strain rate in the flame is driven by the vortex shedding process. Finally, static pressure fluctuation in the cavity shear layer was also observed to be periodic with local maxima occurring at the same frequency as the vortex shedding frequency. Thus the flame strain rate and pressure fluctuations in the cavity shear layer are both driven by the vortex shedding process and are reliable indicators of vortex shedding frequency. 

Three-dimensional simulations were performed with $z$-spans of two and four millimeters using the turbulent inflow boundary condition. In both cases, the flame remained anchored to the cavity leading edge indefinitely with significant lift-off from the bottom wall. The flame surface in the cavity shear layer exhibited considerable variation in the third-dimension. The $Q$-criterion was used to observe vortex structures throughout the combustor, showing larger turbulent structure in the core flow upstream of the flame than downstream, indicating that the turbulence generated at the inflow dissipates to smaller scales as the fuel-air mixture is processed by the flame. Velocity fluctuations showed the turbulent waves generated at the inflow boundary propagate through the entire domain, though the turbulence is strongest upstream of the flame closer to the inflow boundary. The time-averaged turbulence intensity agreed with that observed in the 2D simulations. Future work will include the development of an anisotropic turbulent inflow boundary condition and examine the effect of anisotropic turbulence on the flame structure and stability.


\section*{Acknowledgments}
The authors gratefully acknowledge the support of the Base Program at the Naval Research Laboratory.

\bibliographystyle{elsarticle-num}
\singlespacing
\bibliography{bib_goodwin_titles}

\begin{thebibliography}{10}
\expandafter\ifx\csname url\endcsname\relax
  \def\url#1{\texttt{#1}}\fi
\expandafter\ifx\csname urlprefix\endcsname\relax\def\urlprefix{URL }\fi
\expandafter\ifx\csname href\endcsname\relax
  \def\href#1#2{#2} \def\path#1{#1}\fi

\bibitem{ben2001cavity}
A.~Ben-Yakar, R.~K. Hanson, Cavity flame-holders for ignition and flame
  stabilization in scramjets: an overview, Journal of Propulsion and Power
  17~(4) (2001) 869--877.

\bibitem{rasmussen2005stability}
C.~C. Rasmussen, J.~F. Driscoll, K.-Y. Hsu, J.~M. Donbar, M.~R. Gruber, C.~D.
  Carter, Stability limits of cavity-stabilized flames in supersonic flow,
  Proc. Combust. Inst. 30~(2) (2005) 2825--2833.

\bibitem{micka2009combustion}
D.~J. Micka, J.~F. Driscoll, Combustion characteristics of a dual-mode scramjet
  combustor with cavity flameholder, Proc. Combust. Inst. 32~(2) (2009)
  2397--2404.

\bibitem{wang2013combustion}
H.~Wang, Z.~Wang, M.~Sun, N.~Qin, Combustion characteristics in a supersonic
  combustor with hydrogen injection upstream of cavity flameholder, Proc.
  Combust. Inst. 34~(2) (2013) 2073--2082.

\bibitem{liu2019cavity}
Q.~Liu, D.~Baccarella, W.~Landsberg, A.~Veeraragavan, T.~Lee, Cavity
  flameholding in an optical axisymmetric scramjet in mach 4.5 flows, Proc.
  Combust. Inst. 37~(3) (2019) 3733--3740.

\bibitem{potturi2015large}
A.~S. Potturi, J.~R. Edwards, Large-eddy/{R}eynolds-averaged navier--stokes
  simulation of cavity-stabilized ethylene combustion, Combust. Flame 162~(4)
  (2015) 1176--1192.

\bibitem{rockwell2017development}
R.~D. Rockwell, C.~P. Goyne, H.~Chelliah, J.~C. McDaniel, B.~E. Rice, J.~R.
  Edwards, L.~M. Cantu, E.~C. Gallo, A.~D. Cutler, P.~M. Danehy, Development of
  a premixed combustion capability for dual-mode scramjet experiments, Journal
  of Propulsion and Power 34~(2) (2018) 438--448.

\bibitem{allison2017investigation}
P.~M. Allison, K.~Frederickson, J.~W. Kirik, R.~D. Rockwell, W.~R. Lempert,
  J.~A. Sutton, Investigation of supersonic combustion dynamics via 50 k{H}z
  {CH}* chemiluminescence imaging, Proc. Combust. Inst. 36~(2) (2017)
  2849--2856.

\bibitem{geipel2017high}
C.~M. Geipel, R.~Rockwell, H.~Chelliah, A.~D. Cutler, C.~Spelker, Z.~Hashem,
  P.~M. Danehy, in: 33rd AIAA Aerodynamic Measurement Technology and Ground
  Testing Conference, 2017.

\bibitem{geipel2020AIAAjournal}
C.~M. Geipel, A.~D. Cutler, R.~D. Rockwell, H.~K. Chelliah, Characterization of
  flame front structure in a dual-mode scramjet combustor with oh-plif, AIAA J.
  (in revision) (2020).

\bibitem{peters1999turbulent}
N.~Peters, The turbulent burning velocity for large-scale and small-scale
  turbulence, Journal of Fluid mechanics 384 (1999) 107--132.

\bibitem{nielsen2020AIAAjournal}
T.~Nielsen, J.~R. Edwards, H.~K. Chelliah, D.~A. Lieber, C.~M. Geipel, C.~P.
  Goyne, R.~D. Rockwell, A.~D. Cutler, Hybrid les/rans analysis of a premixed
  ethylene-fueled dual-mode scramjet combustor: Scaled-down cavity
  configuration, AIAA J. (in press) (2020).

\bibitem{dhamankar2018overview}
N.~S. Dhamankar, G.~A. Blaisdell, A.~S. Lyrintzis, Overview of turbulent inflow
  boundary conditions for large-eddy simulations, Aiaa Journal 56~(4) (2018)
  1317--1334.

\bibitem{Davidson07usingisotropic}
L.~Davidson, Using isotropic synthetic fluctuations as inlet boundary
  conditions for unsteady simulations, Advances and Applications in Fluid
  Mechanics 1~(1) (2007) 1--35.

\bibitem{shur2014synthetic}
M.~L. Shur, P.~R. Spalart, M.~K. Strelets, A.~K. Travin, Synthetic turbulence
  generators for rans-les interfaces in zonal simulations of aerodynamic and
  aeroacoustic problems, Flow, turbulence and combustion 93~(1) (2014) 63--92.

\bibitem{schlatter2009turbulent}
P.~Schlatter, R.~{\"O}rl{\"u}, Q.~Li, G.~Brethouwer, J.~H. Fransson, A.~V.
  Johansson, P.~H. Alfredsson, D.~S. Henningson, Turbulent boundary layers up
  to re $\theta$= 2500 studied through simulation and experiment, Physics of
  fluids 21~(5) (2009) 051702.

\bibitem{rauch2018dns}
A.~H. Rauch, A.~Konduri, J.~Chen, H.~Kolla, H.~K. Chelliah, Dns investigation
  of cavity stabilized premixed turbulent ethylene-air flame, in: 2018 AIAA
  Aerospace Sciences Meeting, 2018, p. 1674.

\bibitem{lund1998generation}
T.~S. Lund, X.~Wu, K.~D. Squires, Generation of turbulent inflow data for
  spatially-developing boundary layer simulations, Journal of computational
  physics 140~(2) (1998) 233--258.

\bibitem{Joh20JCP}
R.~F. Johnson, A.~D. Kercher, A conservative discontinuous {G}alerkin
  discretization for chemically reacting navier stokes equations, J. Comput.
  Phys. 423 (2020).

\bibitem{Wil50}
C.~R.~Wilke, A viscosity equation for gas mixtures, J.~Chem.~Phys 18 (1950)
  517--519.

\bibitem{Mat67}
S.~{Mathur}, P.~K. {Tondon}, S.~C. {Saxena}, Thermal conductivity of binary,
  ternary and quaternary mixtures of rare gases, Molecular Physics 12 (1967)
  569--579.

\bibitem{Kee89}
R.~J. Kee, J.~A. Miller, G.~H. Evans, G.~Dixon-Lewis, A computational model of
  the structure and extinction of strained, opposed flow, premixed methane-air
  flames, Proc. Combust. Inst. 22~(1) (1989) 1479--1494.

\bibitem{Got01}
S.~Gottlieb, C.~Shu, E.~Tadmor, Strong stability-preserving high-order time
  discretization methods, SIAM Review 43~(1) (2001) 89--112.

\bibitem{dong2008numerical}
G.~Dong, B.~Fan, J.~Ye, Numerical investigation of ethylene flame bubble
  instability induced by shock waves, Shock Waves 17~(6) (2008) 409--419.

\bibitem{moura2017eddy}
R.~C. Moura, G.~Mengaldo, J.~Peir{\'o}, S.~J. Sherwin, On the eddy-resolving
  capability of high-order discontinuous galerkin approaches to implicit
  les/under-resolved dns of euler turbulence, Journal of Computational Physics
  330 (2017) 615--623.

\bibitem{gassner2013accuracy}
G.~J. Gassner, A.~D. Beck, On the accuracy of high-order discretizations for
  underresolved turbulence simulations, Theoretical and Computational Fluid
  Dynamics 27~(3-4) (2013) 221--237.

\bibitem{Cantera}
D.~G. Goodwin, R.~Speth, H.~Moffat, B.~Weber, Cantera: An object-oriented
  software toolkit for chemical kinetics, thermodynamics, and transport
  processes, version. 2.4. 0, 2018, URL: https://www. cantera. org.

\bibitem{Fia14}
T.~Fiala, T.~Sattelmayer, Nonpremixed counterflow flames: scaling rules for
  batch simulations, Journal of Combustion 2014 (2014).

\bibitem{Che91}
H.~Chelliah, C.~K. Law, T.~Ueda, M.~Smooke, F.~Williams, An experimental and
  theoretical investigation of the dilution, pressure and flow-field effects on
  the extinction condition of methane-air-nitrogen diffusion flames, in:
  Symposium (International) on Combustion, Vol.~23, Elsevier, 1991, pp.
  503--511.

\bibitem{poinsotBook2005}
T.~Poinsot, D.~Veynante, Theoretical and numerical combustion, 2nd Edition,
  Edwards RT Inc., Philadelphia, 2005.

\bibitem{candel1990flame}
S.~M. Candel, T.~J. Poinsot, Flame stretch and the balance equation for the
  flame area, Combustion Science and Technology 70~(1-3) (1990) 1--15.

\bibitem{kolavr2007vortex}
V.~Kol{\'a}{\v{r}}, Vortex identification: New requirements and limitations,
  International journal of heat and fluid flow 28~(4) (2007) 638--652.

\end{thebibliography}
%
%
%
%
%
%
%
%
%
%
%
%
%
%
%
%
%
%

\end{document}